# The extensive photo response on metal/n-Si clarified by the zero-gap with inter-band phonon scatterings

Kazuya Nakayama[a] and Takanari Yasui[b]

[a] *College of Engineering, Chubu University, Kasugai, Aichi, 487-8501, Japan.*

[b] *Independent Researcher, formerly of Nagaoka University of Technology, Nagaoka, Niigata, 940-21 37,*
*Japan.*

Author e-mail addresses: [a] nakayama.kazuya@fsc.chubu.ac.jp, [b] apxj91017@gmail.com

**Abstract**

UVA to NIR with multi-directional photo responses have been found on metal (Au)/n-Si device. A reasonable explanation has not been found in various physical models of Si-devices for the phenomena. We approached a zero-gap at X (reciprocal point) in two conduction bands of Si to analysis the optical response with the inter-band phonon scatterings.

The calculation of the quantum efficiency between X-Γ and X-W successfully simulated the sensitivities in visible region (1.1 to 2.0 eV), the carrier density profile well fitted the response in NIR (0.6 to 1.0 eV). Filling up the zero-gap by doping electrons (~10^18/cm$^3$) at around X, a lower limit of 0.6 eV arose in the measurement below Si-band gap of 1.17 eV.

Indirect/direct transitions of inter conduction bands: X-W, X-K and Γ-L in the 1st Brillouin Zone/Van Hove singularity at L point, synchronizing with phonon scattering, gave a variety of directional photo-responses. The carrier scattering model for the inter bands (X-W, X-K and Γ-L) were consistent with the directional dependence of photo-currents under UVA (3.4 eV) and Visible (3.06 eV) excitations. Band to band scatterings assisted to extend the available optical range and increase its variety of directional responses. Utilizing this principle, a new frontier will be opened in the photo-conversion system by using indirect-transition semiconductors and thus, it will be released from those band gaps and directivity limitations.

## Introduction

We already reported the super wide: 400 to 2200 nm optical-response, being satisfactory in practical efficiency, by the metal (Au) on n-Si device as shown in Fig. 1.[1,2,3] The light source (50 W tungsten lamp) was monochromatized by a conventional spectrometer (f=50 cm) with a resolution of ~1.5 nm. The excitation power was calibrated by a conventional p-i-n photo diode for the range to 1100 nm. A pulsed-laser excitation (the peak power of 0.1 mJ, pulse width of 20 ps) was used for more than 1100 nm. Following this, the photo-current was amplified by a factor of 10 through an I-V amplifier and transmitted to a digital real-time oscilloscope (Tektronix TDS 620B; bandwidth 500 MHz; sampling rate 2.5 GS/s). The latter excitation power was deduced from the wave profile monitored by the oscilloscope, thus, the errors of sensitivity resulted in relatively larger than those for less than 1100 nm. The sensitivity was converted to quantum efficiency (Q. E.) normalized by the excitation photon energy as shown in Fig. 1.

Au electrodes (~ 2 mm$^2$, in Fig. 2 (a)) were directly formed on the Co (8 nm thick)/n-Si surface which was pre-annealed at 600°C as reported in Ref. 1. The CoSix/Si formed by the annealing promoted the conductance of the in-plane surface and through electrodes. The built-in, self-organized, crystal faces of the metal (Au) on Si (100) were observed in SEM image as shown in Fig. 2 (b).[3] These characteristic structures were mostly found in the photo-reactive area, ~3 mm apart from the anode, thus we selected an excitation position in this study (see Fig. 2 (a)).

The extensive optical-responses, however, have not been given any reasonable interpretations for: the response lower than Si-band gap at 1.17 eV (1060 nm), UV response at 365 nm (3.40 eV), multi-directional photo responses (presented in this study) and photo-currents enhanced at the metal edges (to be opened elsewhere).

On the other hand, the combination of metal/Si with this sample has led us to apply Schottky model[4] for the results in the early studies. S. M. Sze reported the photo carriers generated at the depletion layer in Schottky interface and showed a following relation derived from Fowler's theory[7]:

$$\sqrt{R} \sim h(\nu - \nu_0) \quad 1.$$

where hν is the incident photon energy in eV, $R$ the photo-sensitivity in mA/W and the $h\nu_0$ the barrier height estimated from an extrapolated value at the photon energy axis. Fig. 3 shows a barrier height of 0.45 eV for our sample much differs from 0.8 eV[8] formerly reported for the same combination of Au/n-Si when using Schottky model. Furthermore, noticeable responses less than 1.17 eV (1060 nm) of Si-band gap means Schottky model cannot be applied to our original devices.

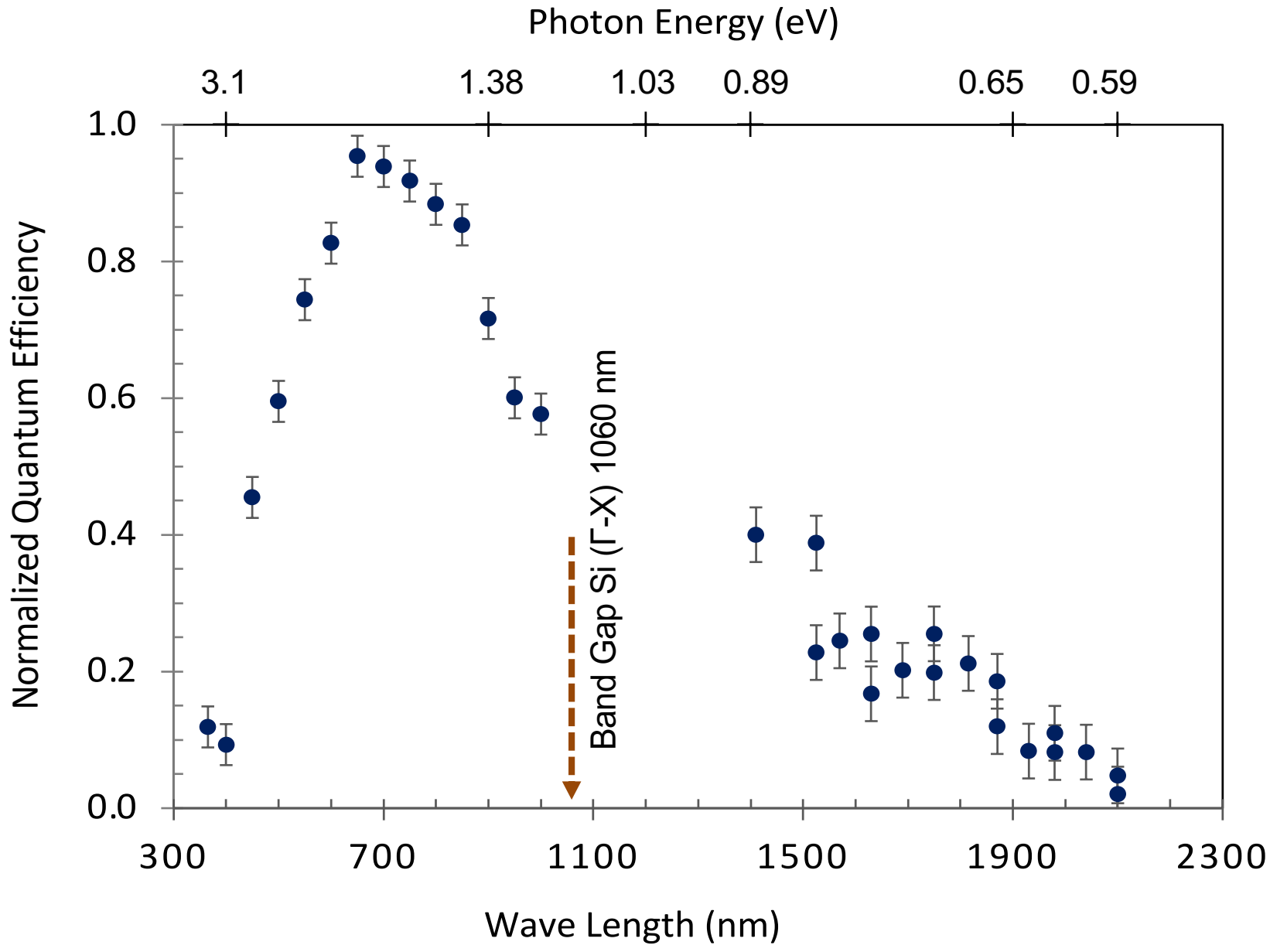

Fig. 1 Available operation of 350 nm (UV) to 2100 nm (NIR) released from Si band gap limitation.[2] An anomalous decline at ~900 nm suggested the multiplex processes to generate photo-carriers. The sensitivity peaked at 600 nm, corresponding its quantum efficiency of 95%. A variation in error bars due to the different light source: tungsten lump for the range up to 1100 nm and pulsed laser for another range.

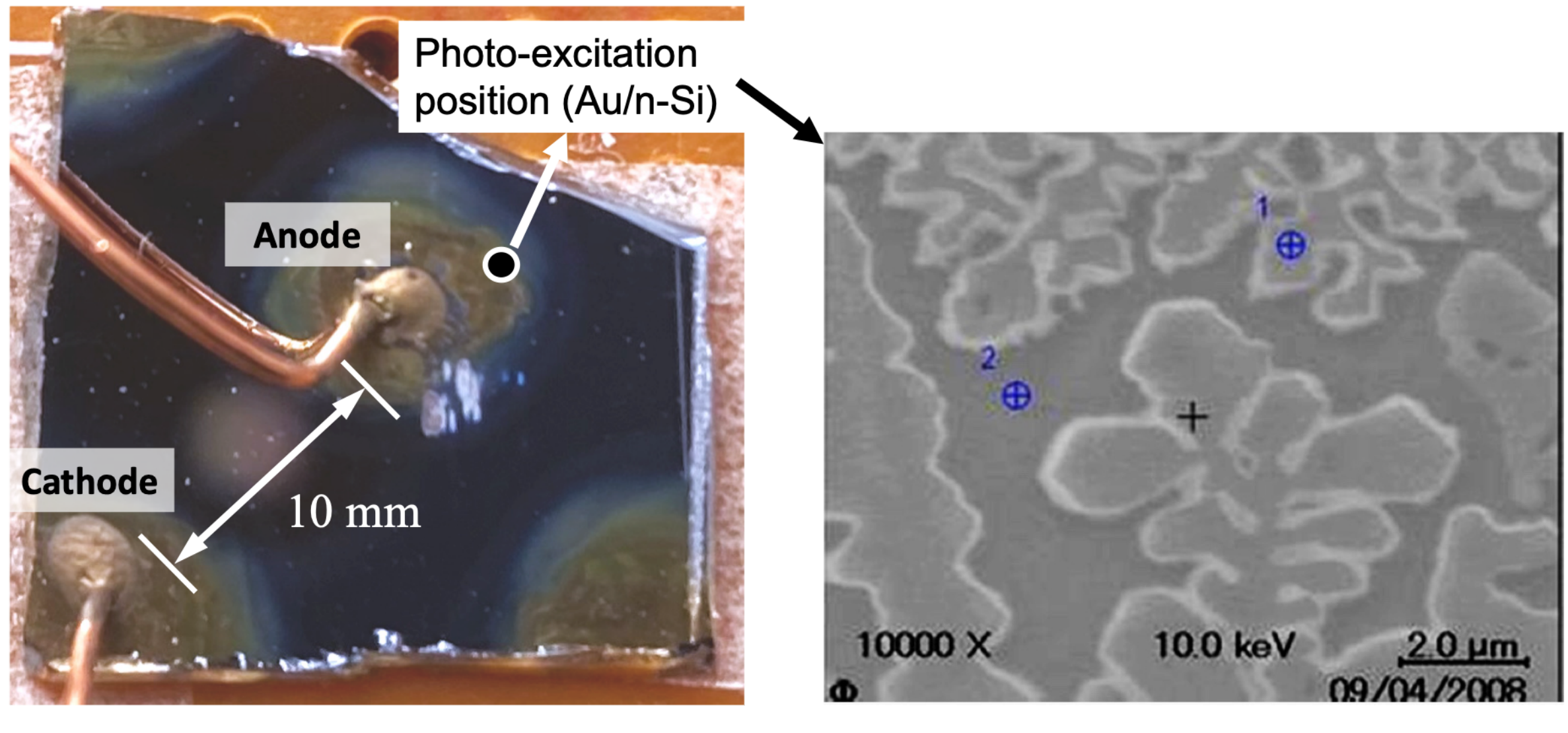

Fig. 2 (a)

Fig. 2 (b)

Fig. 2: Images of device structures. (a) The photo-excitation position (a closed circle) at ~3 mm distance from the anode. (b) SEM image of surface structures on the photo-excitation area showed the self-organized structures (Au) on n-Si.

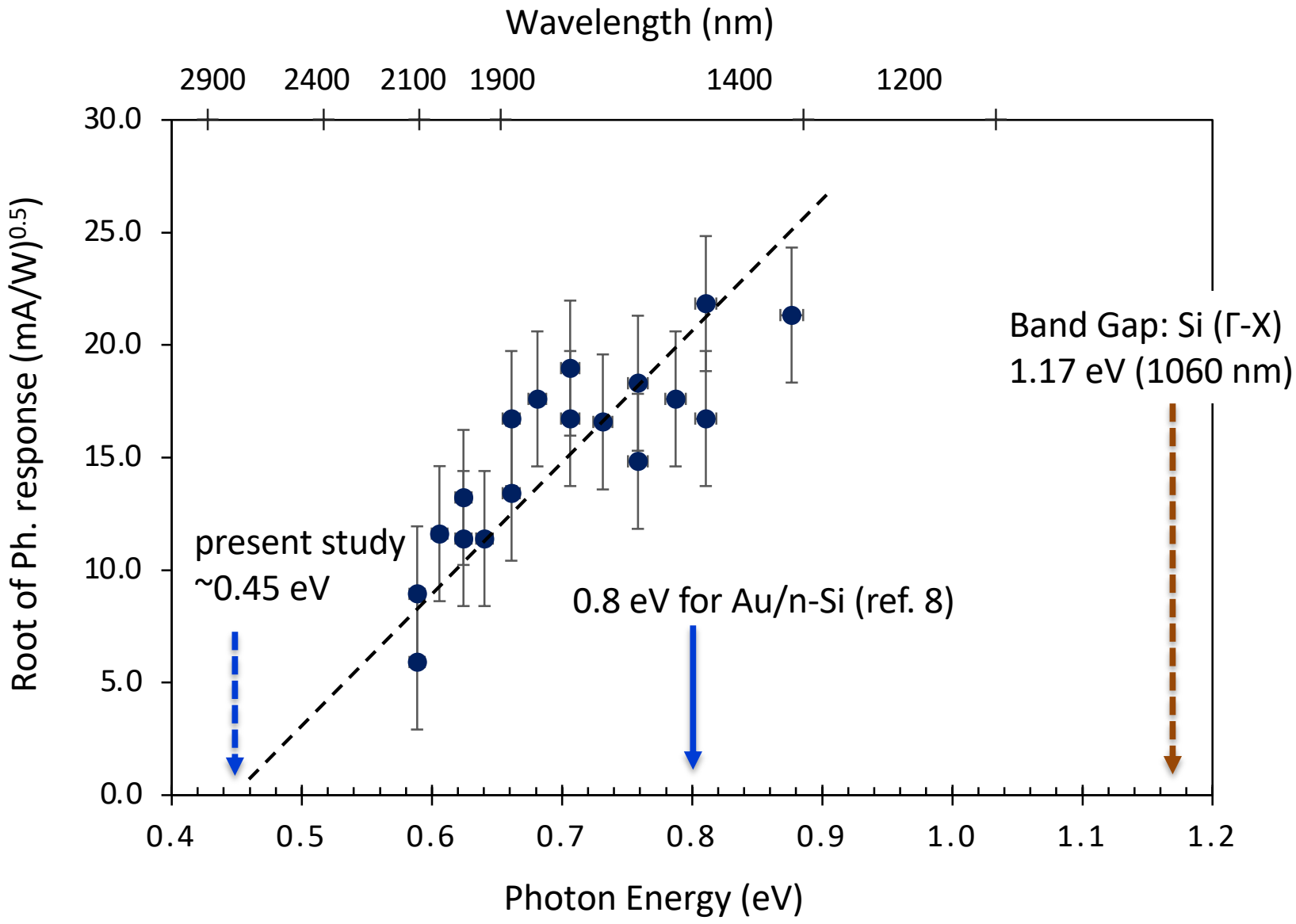

Fig. 3 The photo response $(mA/W)^{0.5}$ and incident photon energy (eV) shows a linear relation for the Schottky barriers.[7] A barrier height of 0.45 eV extrapolated in this study not agreed with 0.8 eV previously reported for Au/n-Si.[8]

On the other hand, multi-directional photo response was confirmed on top, reverse and diagonal to the sample surface as shown in Fig. 4. The conventional tungsten light source was focused on the sample surface with the irradiation area ~3 $mm^2$ for this measurements. The reverse operation through Si substrate was due to the higher transparency in the NIR (> 1000 nm) region of Si[9] and the tungsten light source including its wavelength range of visible to NIR.

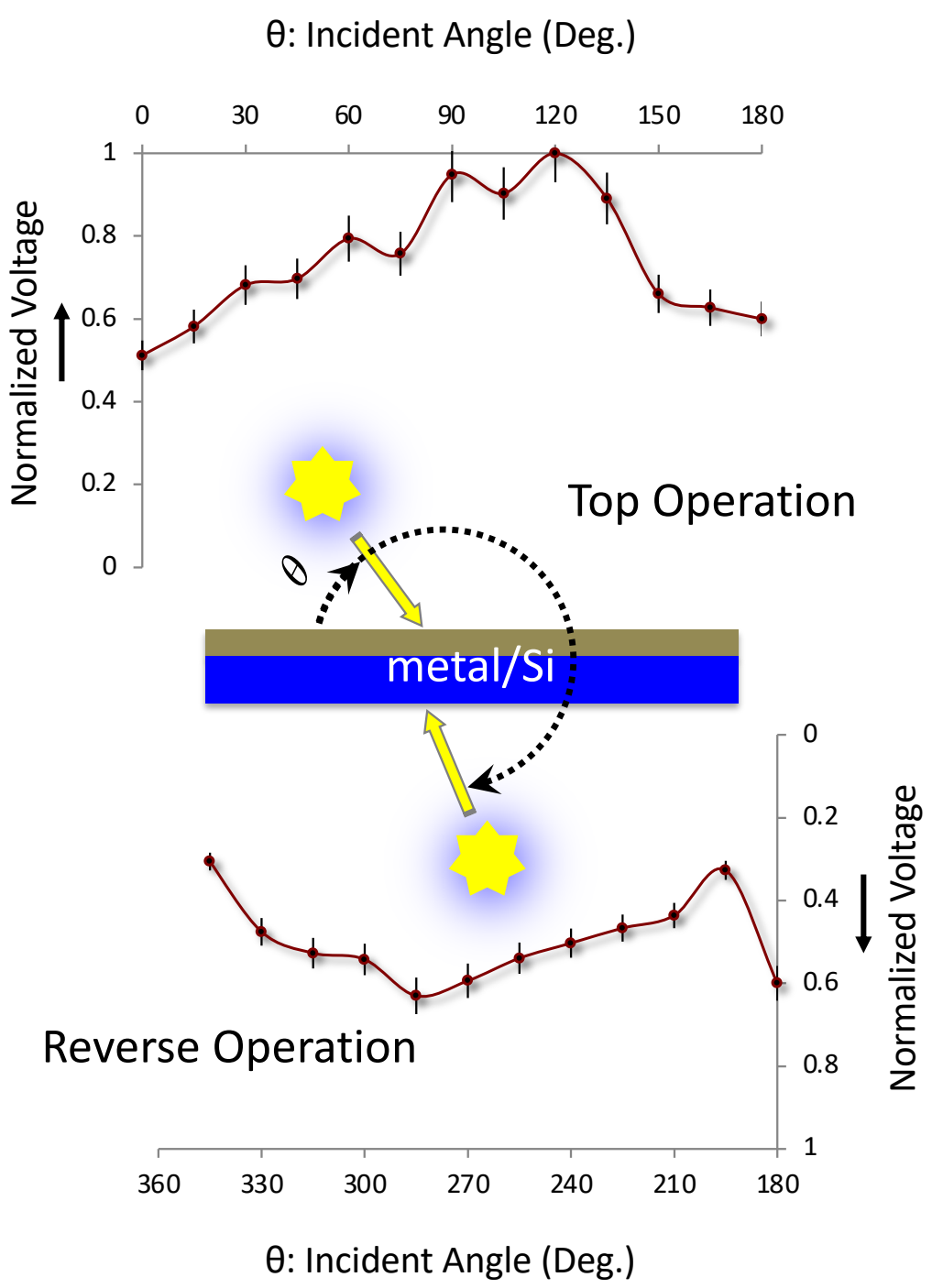

Fig. 4 Top and reverse operations and ~50% decrease for in-plane irradiation at *q*=0°, 180°.

These results above suggested the renewal model to be applied for the phenomena found on metal crystals/n-Si which are different from the characters ever reported on Si-related photo conversion systems.[4,5,6]

We approached the zero gap at around a reciprocal lattice point X with wave number (*k*) of 2.0 in unit of π/a $cm^{-1}$, where “a” is the lattice constant of Si (= $5.431\times10^{-8}$ cm, see Fig. 5). The conduction bands from X-Γ (X1), X-W(K) (Xu) consist a zero gap at around X, *k* ~ 2.0, where X, W and K are the reciprocal lattice points in 1st Brillouin Zone of Si (see Fig. 6).[10,11,12] We proposed the indirect transitions within these conduction bands (process “III” introduced in Sec. 2-1) because the zero gap was expected to realize NIR response below the Si band gap.

Another possibility of carrier generation is the direct transition at Van Hove singularity of L point in Si[13,14,15] (process “I” in Sec. 2-1) because the response observed for UVA region as shown in Fig. 1.

We calculated quantum efficiency of I and III obtained by optical transition probability under stable oscillator strength as mentioned in later Part. 2. The doping model (process “IV” in Sec. 2-1) will give a suitable approach to the responses lower than Si band gap because some carriers will be filled in the bottom of conduction bands (X1 and Xu) at zero gap (*k*~2.0).

Multi-directional scanning is required when any effects from 3D direction, e.g., from Lc (Γ-L direction), are independent on in-plane carrier transitions between X1 and Xu. We introduced the modified scattering model for the simulation with respect to the sample rotation angles ($\phi$, $\theta$), incident photon direction and excitation wavelengths. It will be expected to confirm if the cooperation occurs within some conduction bands in Si, including direct transition (I) and indirect transitions (III).

This study clarified the origin of those features for metal/n-Si crystal system in terms of calculation and experiment.

## 1 The surface structure of metal/n-Si and those function

Crystal faces have been observed on the metal (Au) surfaces as shown in SEM image in Fig. 2 (b).[3] The thicknesses of metal are several 10 nm, in-plane width within sub-micron and the aspect ratio of width/thickness is 10 to 100 which were confirmed by AFM measurement. The photoreactive area corresponded with the surface where these characteristic structures have been observed. The metal structures were self-organized by thermal annealing processes as previously reported[1] resulted in variety of crystal surface directions.
It is supposed that the metal structures exchange incident photon for surface plasmon[16] and give rise to variety of directions when it functioning as optical waveguide into Si surface and led to multi-directional operation of our Au/n-Si device.
Previously, the surface plasmon improved performance of photo-diodes[17] and solar cells.[18] In contrast to their effects, wide optical response and multi directional operations were unique and advanced features demonstrated by our metal/n-Si device.

## 2 Quantum efficiency and photosensitivity

### 2-1 Photo absorption processes and band structure of Si

Considering the unique phenomena observed on metal/n-Si, we calculated quantum efficiency of the photo-currents, supposing the induced carriers were originated in Si crystal. Fig. 5 shows energy band diagram of Si crystals with energy (eV) in vertical axis on top figure and those energy gaps in the bottom.
The possible carrier generation processes are listed as follows:
0: Γ-X indirect transition (except for calculations in this study)
I: Van Hove singularity at reciprocal lattice point L[10,11,12]
II: Van Hove singularity at reciprocal lattice point X[10,11,12]
III: X1-Xu indirect transition
IV: n-type carrier doping[19]
Doped carriers exist preliminarily in the bottom of X1 as well as Xu in this study. This means Γ-X indirect transition is a lower priority for the photo-excitation processes, besides that, the lowest energy gap at Γ-X is exchanged for the minimum gap between X1 and Xu (see Sec. 2-4). This is why the process 0 was excluded from the calculations.
UV response by process I and II contribute photo carrier generation in the conduction bands: Lc and Xc mostly through each Van Hove singularities. Process II has been eliminated from the calculations since those energy gaps (3.45-4.27 eV) are larger than maximum excitation photon energy of 3.40 eV in this study. Although the carrier transfer is occurred through phonon scatterings, calculation in here is consisted of the transition probability (~coupling density of state: $DOS_{CV}$) and the mobility. The initial requirement of n-carriers in the lower band (X1) for the process III is satisfied with the process IV. The representative model of carrier doping[19] was referred for the calculation of carrier density in the conduction band. The calculated efficiencies were evaluated in comparison with the quantum efficiency measured as a function of incident photon energy (eV) in Part 2-2. The inter-band phonon scatterings will be evaluated in the later Part 3.

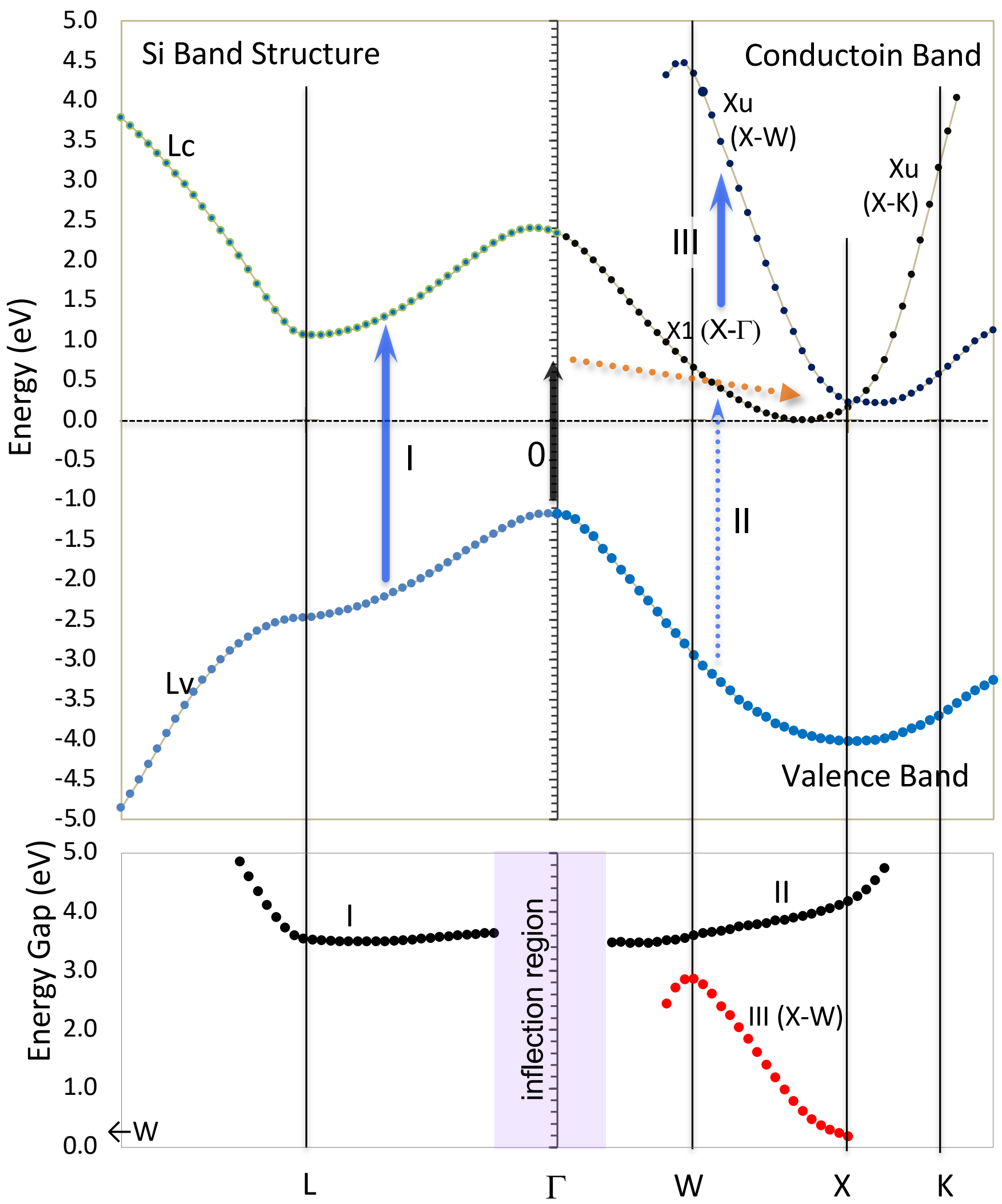

Fig. 5 Photo absorption processes with those notations (I to III) in Si Band Structure (top) and corresponding energy gaps (bottom). Photo-excited carriers relaxed into the bottom of conduction band at X through phonon scatterings (a broken arrow through X1). Some of effective masses are divergent at those inflection points in the band structure. This is why the shaded region was excluded from the calculations. A zero gap in the process III exists at X, X1 across Xu at there, where X1 is the lower conduction band in the direction of X-Γ, Xu the higher one with X-W(K) direction.

Table 1 The selected combinations of *k* and Eg which will be used for those interpretations in Sec. 3-2A and Sec. 3-2B.

| Carrier Generation Process | *k* (x2π/a $cm^{-1}$) | Energy Gap (eV) | Related Lower-Higher Band |
|---|---|---|---|
| I | 1.313 | 3.494 | Lv-Lc |
| III | 0.801 | 2.849 | X1-XW |
| III | 0.801 | 4.068 | X1-XK |
| III | 0.620 | 2.120 | X1-XW |

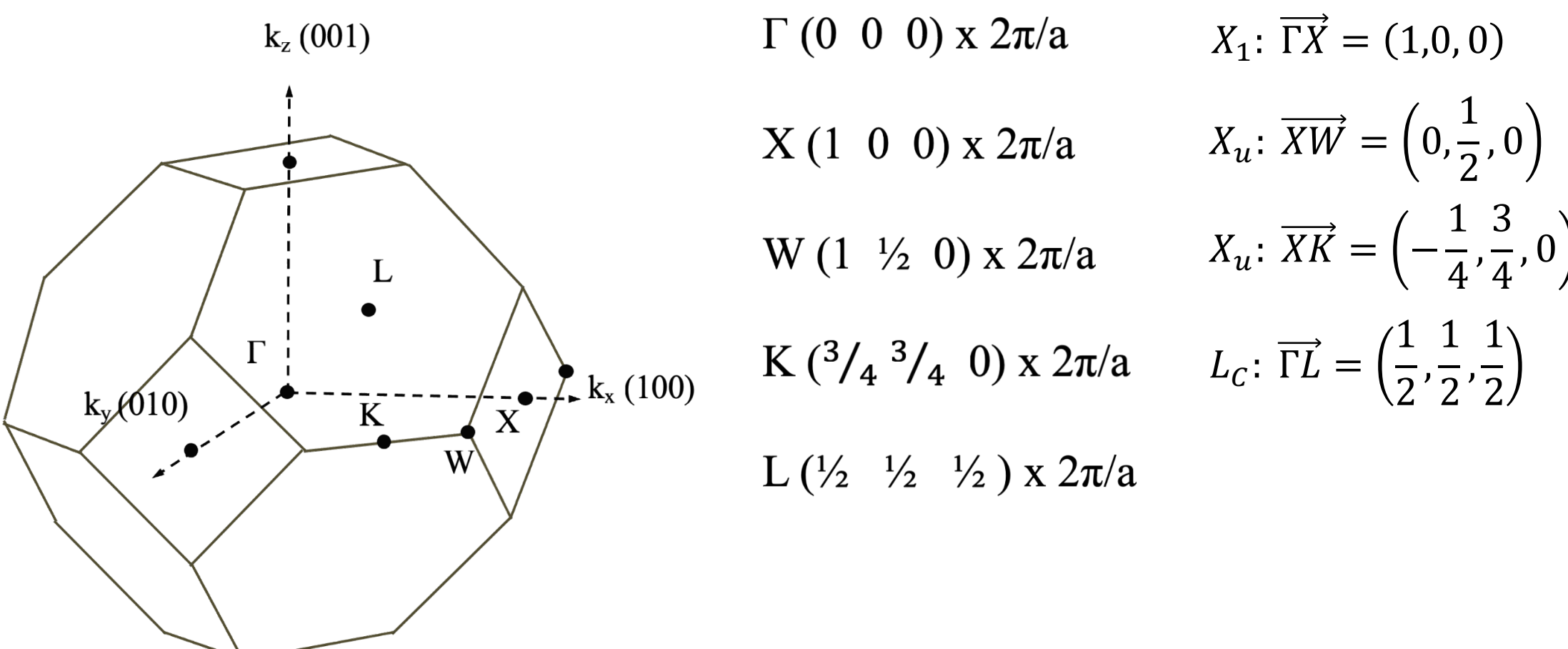

Fig. 6 Related Reciprocal Vectors in 1st Brillouin Zone and those components of reciprocal lattice points, ibid. vectors. Components are in unit of 2π/a $cm^{-1}$ where "a" is the lattice constant of Si.

## 2-2 Calculation processes and quantum efficiency

Following the fundamental theory of photo absorption coefficient ($\alpha$)[20,21], oscillator strength (M) and $DOS_{CV}$ are required for the simulation, as expressed by

$$\alpha \propto M^2 \times DOS_{\mathrm{CV}} \tag{2.}$$

where $$M = |\langle k_C|A\cdot p|k_V\rangle|,\ \ DOS_{\mathrm{CV}} = \Delta(E_C(k) - E_V(k') - \hbar\omega) \tag{3.}$$

We supposed the constant oscillator strength: M when the atomic periodicities (potential variations) can be ignored in the related energy bands. Thus, transition probability can be obtained from the other part in Eq. 2 (Coupling density of state ($DOS_{CV}$)):

$$DOS_{\mathrm{CV}} = 2\left(\frac{1}{2\pi}\right)^3 \int_S \frac{dS}{\left|\frac{dE_C(k)}{dk} - \frac{dE_V(k)}{dk}\right|} \tag{4.}$$

as a function of $k$. The density of state ($DOS_{CV}$) here is calculated when the upper (conduction) and lower (valence) band related to each other through the photo excitation. $\alpha$ in Eq. 2 is proportional to photo carrier density ($n$). Thus, $DOS_{cv}$ gives $n$ as one of factors in the observed photo current:

$$J_{CV} = -\mu \cdot n \cdot eE \tag{5.}$$

where $\mu$ is the mobility, $n$ the carrier density, and $E$ the applied electric field and $\mu$ is given by the next relation:

Mobility (μ)[22,23]:

$$\mu = \frac{1}{\frac{1}{a}\left(\frac{m^*}{m_0}\right)^{\frac{5}{2}}(T)^{\frac{3}{2}} + \frac{1}{b}\left(\frac{m^*}{m_0}\right)^{\frac{1}{2}}(T)^{-\frac{3}{2}}} \tag{6.}$$

where m* is the effective mass:

$$m^* = \frac{\hbar^2}{\left(\frac{d^2E(k)}{dk^2}\right)} \tag{7.}$$

Factors: a and b in Eq. 6 were referred values in Ref. 24-25 at the bottom of conduction band around X, under room temperature. The quantum efficiencies can be estimated by $DOS_{cv}$ (Eq. 4) with the effective mass (μ, Eq. 6) under constant electric field $E$. These conditions led the number of photo induced carriers, i.e., photo current (Eq. 5), are in proportion to the $DOS_{CV}$ multiplied by mobility $\mu$.

Quantum efficiency calculated from mobility and $DOS_{CV}$:
The calculated mobilities are shown in Fig. 7 (a) and $DOS_{CV}$ in Fig. 7 (b), respectively. The horizontal axis is same as Fig. 5. The shaded region was excluded from optical transition processes because their effective masses are diverged at some inflection points. Furthermore, the lowest energy gap around X will be given a priority in the optical transitions between X1 and Xu by doping.

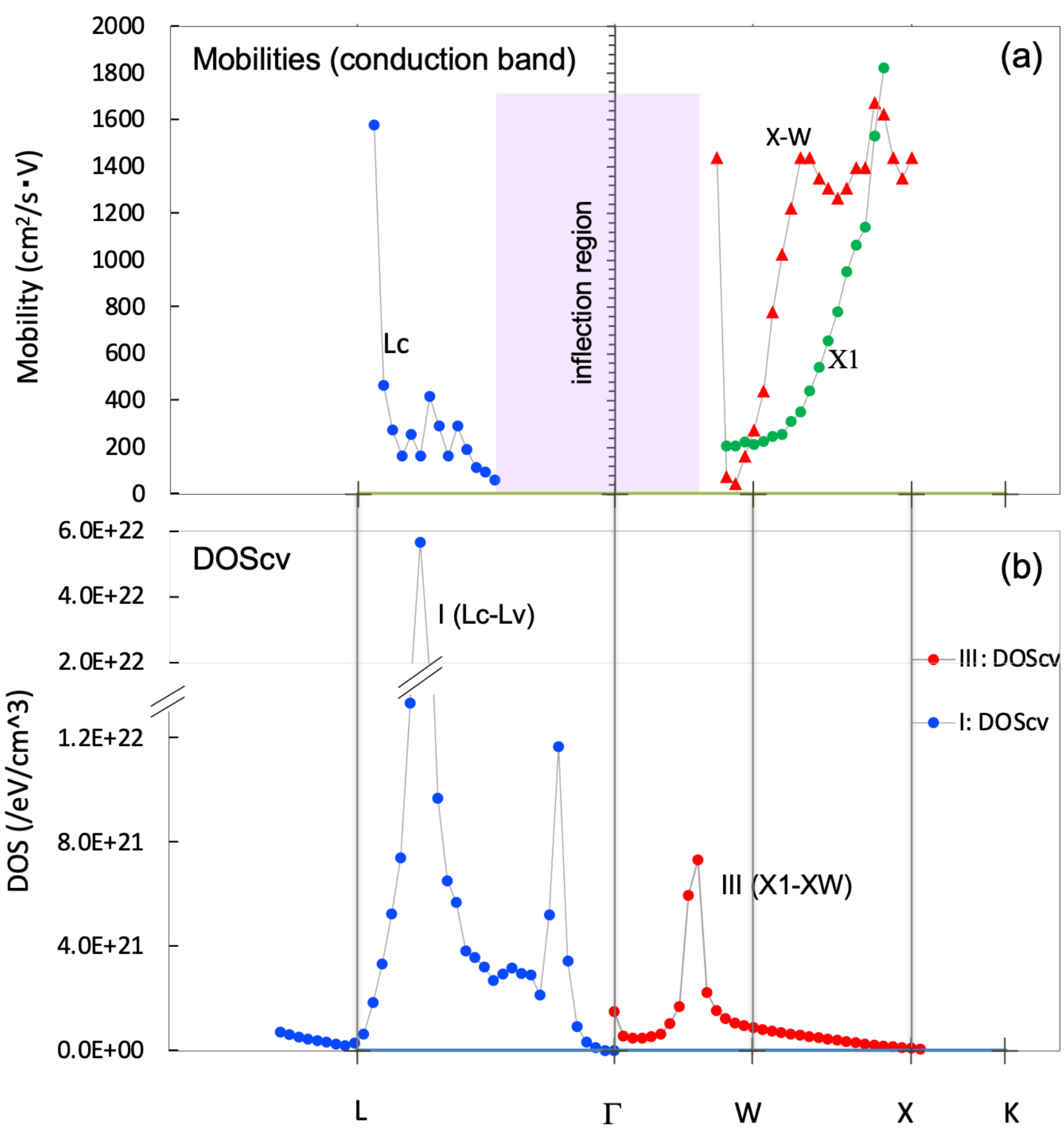

Fig. 7 Calculation results: (a) Mobility as a function of reciprocal points for processes I (Lc) and III (X-W), excepting the inflection points of effective mass. Wave numbers from G (in unit of $\pi$/a cm$^{-1}$): $k$=2.0 at X and $k$=1.73 at L. (b) A maximum $DOS_{CV}$ (5.66x10$^{22}$ /eV/cm$^3$) is confirmed at $k$=1.313 (I: Lc-Lv) (see Table 1).

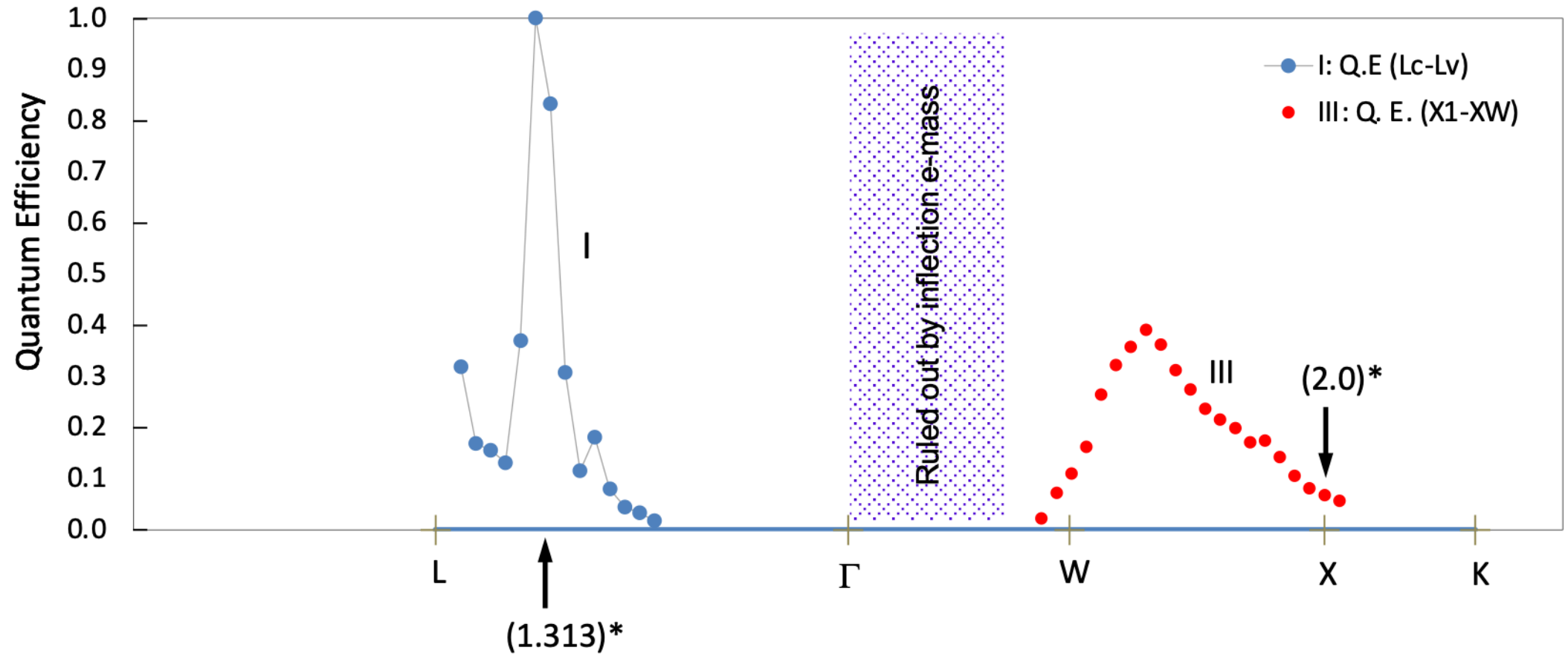

Fig. 8 Calculated quantum efficiency (Q. E.) normalized by a maximum of I for the corresponding points, being proportional to the wavenumber ($k$), plots for process I in the left part and III in the right part, respectively. ( )*: The wave number from Γ in unit of $\pi$/a cm$^{-1}$.

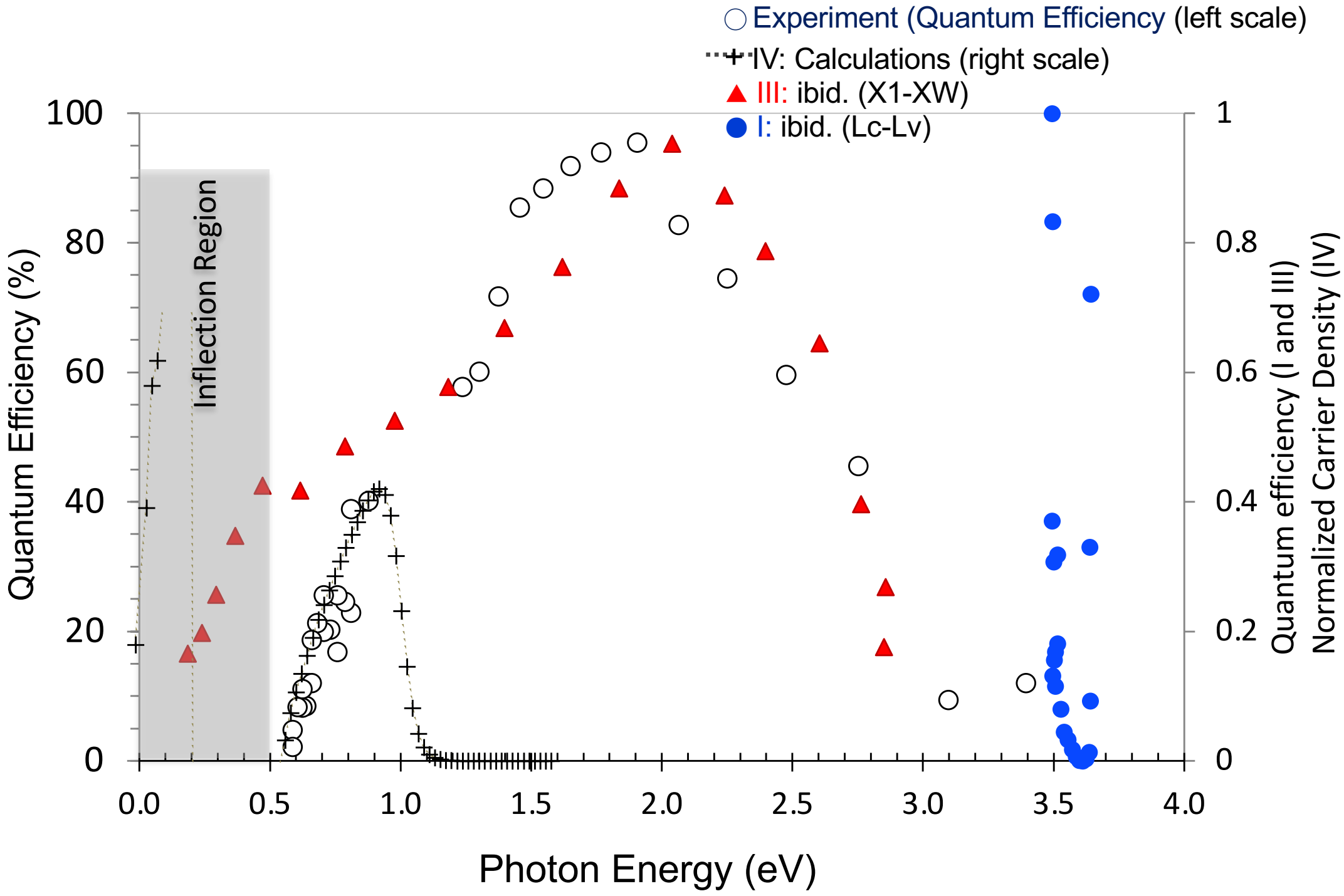

Fig. 9 The quantum efficiency calculated for process I (●) and ibid. III (▲) as a function of the excitation photon energy (eV) corresponding the energy difference of the related bands in Fig. 5. The calculated efficiencies for I (Van Hove singularity) at ~3.5 eV are larger against smaller values in the experiment. The process III closes to measured results within the photon energy of 1.0 to 2.7 eV. Process IV enhanced the photo excitations less than 1.0 eV.

Eg

Fig. 8 shows the quantum efficiency for the corresponding reciprocal points, being equivalent to a function of *k*. Excitation photon energy corresponds with energy difference between lower and upper band in Fig. 5. This leads to assign *k* (horizontal value) to those photon energy (i.e., horizonal value in Fig. 8 and Fig. 9). Quantum efficiency (vertical in Fig. 8) was normalized photo current (Eq. 5) given by DOScv multiplied by effective mass. Following these processes, Fig. 9 shows the quantum efficiency for process I (●) and ibid. III (▲) as a function of the excitation photon energy (eV) in comparison with measured quantum efficiency.

2-3 Optical response in the visible region and the “zero gap” in the inter-band (X1 and Xu): III

Newly introduced process III will be expected to generate photo-induced carriers with visible to NIR excitation region because III includes the visible to a zero gap (see Fig. 5). On the other hand, the phonon scattering is indispensable for the inter-band transition because the direction of X1 differs from that of Xu, being disadvantageous to the optical transitions. Electrons are preliminary not existing at around X without n-doping for Si. Therefore, the zero gap at X in Si crystal has not been paid attention to its usability until now.
Supplying sufficient carriers to the lower band: X1 by the doping (IV), the quantum efficiency between X1 and Xu, without phonon scattering effects, shows useful values with the photon energy less than1.0 eV in Fig. 9. The calculation (+) closely simulates the measured sensitivities (○) within the excitation photon energy of 1.0 to 2.7 eV in Fig. 9. Besides process III, the contribution from process I is supposed to supply (or receive) additional photo carriers. It should be considered that the photo carriers are formed through the inter-band phonon scatterings between these related conduction bands: X1, Xu and Lc. The quantitative analysis through scatterings will be given by the multi-directional analysis in Part 3.

2-4 Doping effects on the NIR photo excitation region: IV

The n-type carrier doping (IV) supplies the bottoms of X1 and Xu with electrons, resulting in the lowest energy gap was given around X in our sample. The carrier density (Nd) in the conduction band was calculated from the DOS multiplied by Fermi-Dirac function.[19] The Nd (●) and DOS (curve) vs band energy are displayed in Fig. 10 (a) for Xu and Fig. 10 (b) for X1, respectively. Subtracting Nd of Xu from ibid. X1, the difference (=ΔNd (X1-Xu) was positive above 0.5 eV as shown in Fig. 10 (c). ΔNd shows the difference up to $6x10^{19}$ ($/cm^3$) in the electron density between X1 and Xu, so it allows X1 to activate as a valence band.
The variation in ΔNd (X1-Xu) (+) in Fig. 10 (c) well simulated the measured sensitivities (○) as a function of photon energy with a doping density at $\sim 10^{18}$ $/cm^3$, when they are normalized by those peek value at ~0.9 eV. The lower limit matches exactly at a photo-excitation energy ~ 0.6 eV simultaneously (Fig. 10 (c)). This is why the doping model (+) gives the better fitting in comparison with the process III (▲) below 1.0 eV as shown in Fig. 9. The doping effects take priority to that of process III for the NIR region (below 1.0 eV, above 1300 nm).
Supposing a minimum energy gap between X1 and Xu, the measured photo sensitivity can be explained by using the doping process: IV in the photo excitation energy of 0.5 to 0.9 eV, below the energy band gap of Si: 1.17 eV (1060 nm).

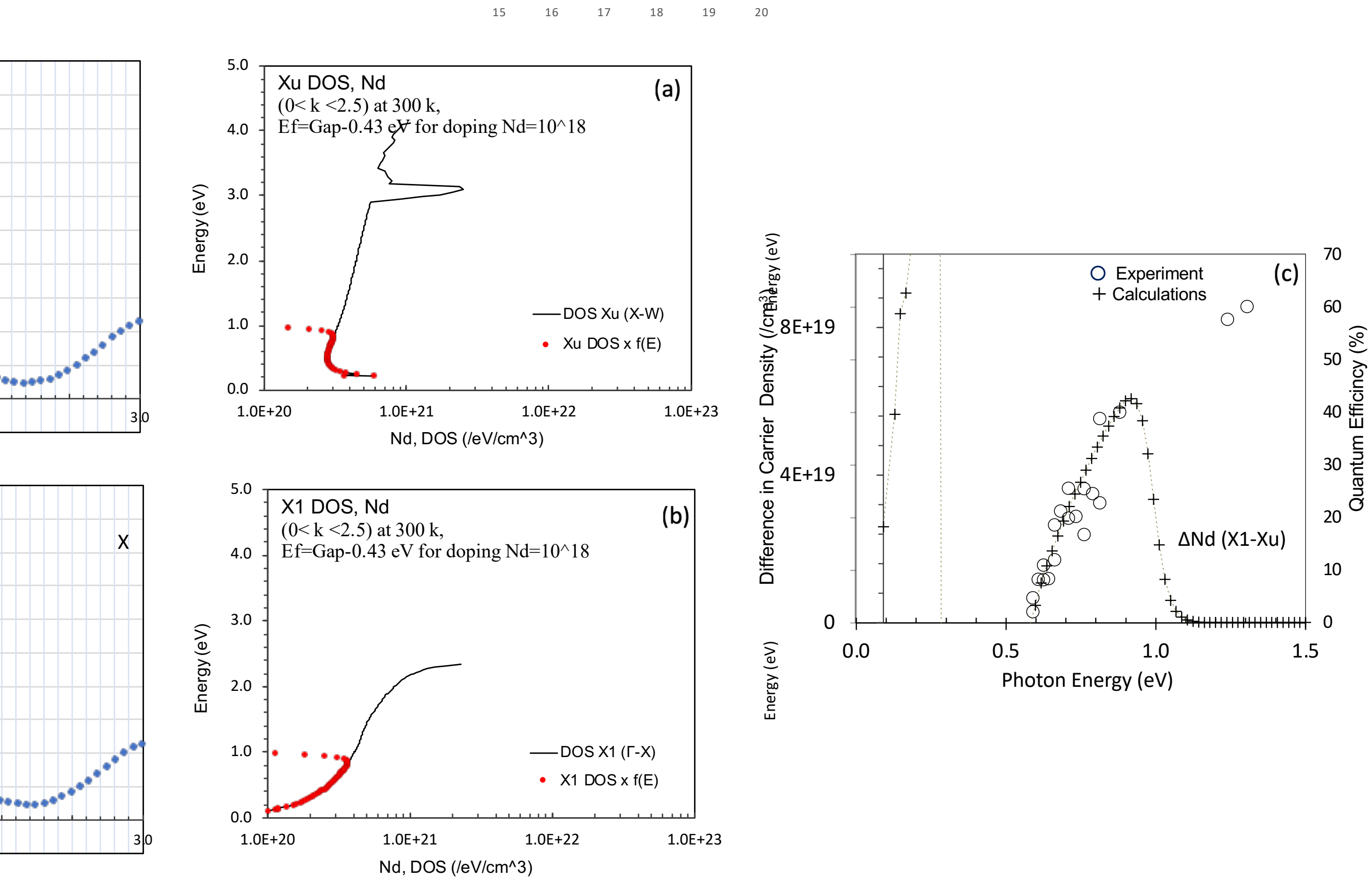

Fig. 10 Calculations of DOS and carrier density profile for $Nd=10^{18}/cm^3$, compared with measured photo sensitivity: (a) DOS (curve) and carrier distribution (closed circles) plots for Xu. (b) ibid. for X1. Those distribution were calculated by DOS×Fermi-Dirac function (see Sec. 2-4). (c) Difference in carrier density (=ΔNd (X1-Xu), the left scale) is closely fitted the quantum efficiency in the right scale as a function of excitation photon energy (crosses with dotted line).

2-5 Van Hove singularity in the UV region: I

A large quantum efficiency was expected by Van Hove singularity at reciprocal lattice point L (process I at $k$~1.3 in the left part of Fig. 8. for the UV absorption around ~3.5 eV in Fig. 9. The estimation for I (●), however, was larger than the observed photo sensitivity (○) because of a mismatch between the measured photon energy at 3.4 eV and an optimum condition at 3.5 eV. In contrast to this, the estimation (▲) closed to the measured results (○) under 1.0 to 2.7 eV excitation in Fig. 9.
The calculations above give the transition probability within the inter band gaps. On the other hands, the wave vector of the incident photon requires to match with the direction of the lower band, being another indispensable factor. This is because all related conduction bands here are originated from various directions in Si crystal as shown in Fig. 6. Phonon scatterings compensate for the mismatch throughout the transition
The multi-directional analysis in the next section is essential in view of the wave number, i.e. "$k$-vector", since the band-to-band scatterings occur within the inter conduction bands: Lc, X1, and Xu corresponding to each of their directions.

## 3 Multi directional analysis of photosensitivity.

The top and reverse operation (Fig. 4) leads to the more precise 3 D spectroscopy in order to confirm the multidirectional scattering within the inter conduction bands: X1 to Xu in $k_x$-$k_y$ plane, Lc to the others in $k_z$-$k_x$ (or -$k_y$) space in the 1st Brillouin Zone (see Fig. 6).
The photocurrents for sample rotation angles ($\phi$ and $\theta$ in Fig. 11) were measured under photo excitation wavelengths: 365 and 405 (nm). The current flow measured in (110) direction which was fixed by two electrodes on the sample surface, $\phi$ ($\theta$) rotation corresponds to the scanning in $k_x$-$k_y$-$k_z$ ($k_x$-$k_y$) plane, respectively. The light source was vertical to the (100) sample surface, excitation area about ~1mm$^2$, the powers below several 10 μW and the relative arrangement as displayed in Fig. 11. The positions of the reciprocal lattice points vary with the angles of $\phi$ and $\theta$ against a fixed direction of the incident photon and a rotation center at X prior to Γ as shown in Fig. 12.
The scanning of $\phi$ with a rotation axis (Γ-X) enables it to investigate the scattering effect on photocurrent between Lc and X1 (or Xu). On the other hand, $\theta$ scanning with a rotation origin at X will be reflected in the photocurrents with scatterings between X-Γ (X1) and X-W(or K) (Xu).

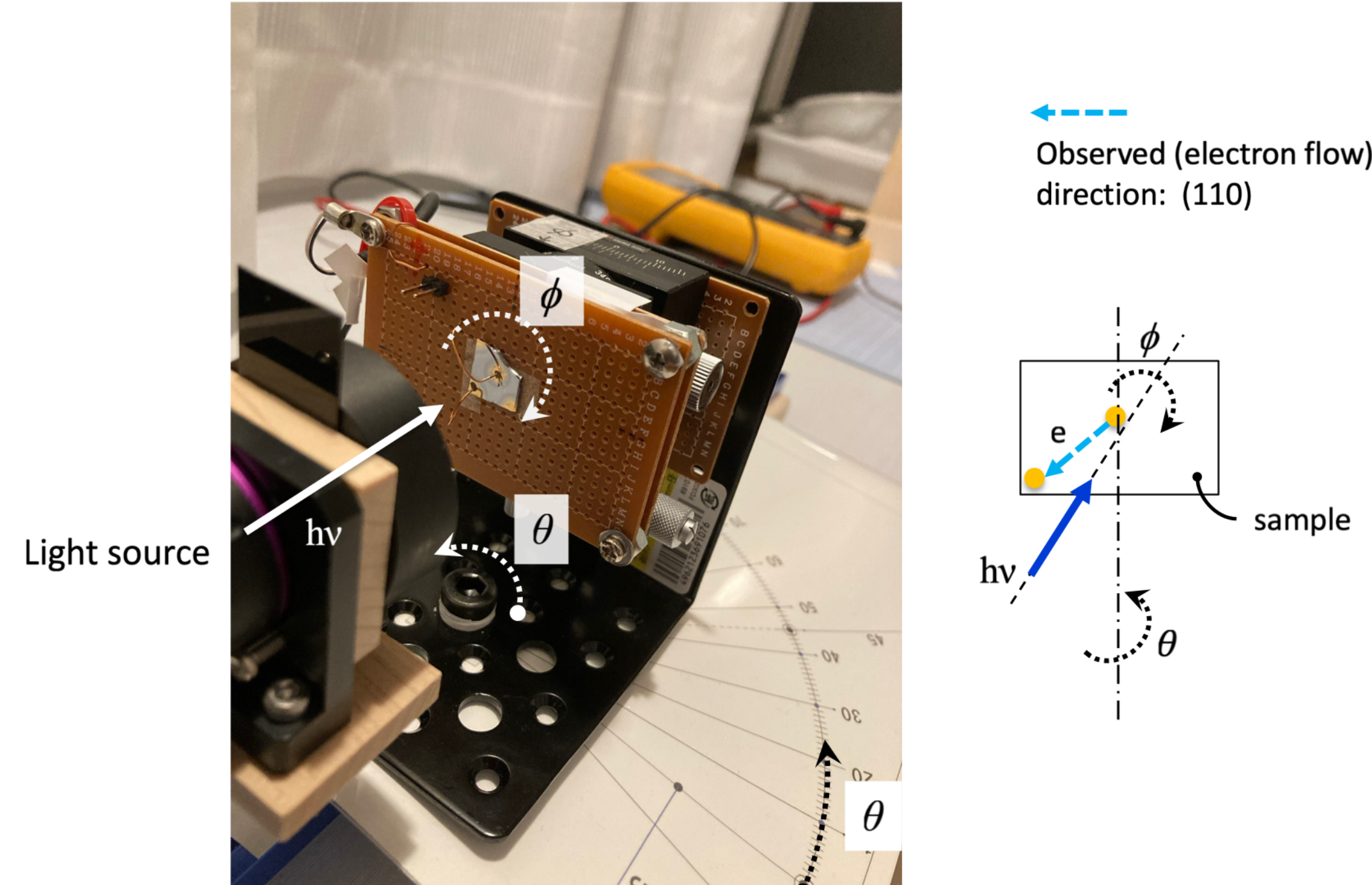

Fig. 11 Close up image of a sample set in the directional ($\phi$, $\theta$) analysis system in the left. An photo-excitation position on the sample surface was indicated in Fig. 2 (a).

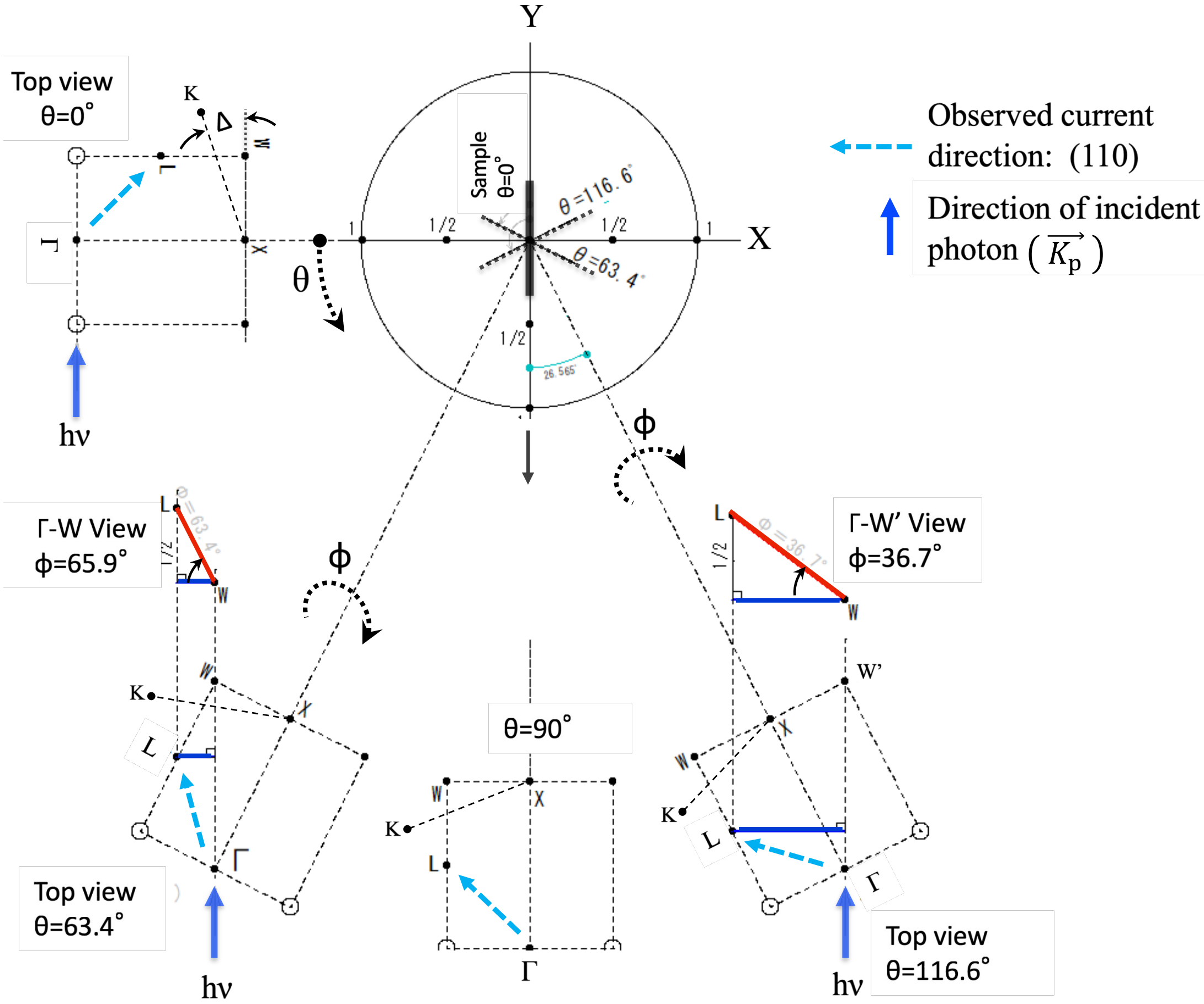

Fig. 12 Schematic diagram of directional analysis: variation in arrangement of reciprocal lattice points by the sample rotation with($\phi$, $\theta$), measured in (110) direction and an incident photon in parallel with X coordinate. Characteristic excitation is realized at $\theta$=63° and 117° because Γ-W is arranged in parallel with incident photon. Crossing angles of inter-vectors (vector components in Fig. 6): Δ=18.44° for $\overrightarrow{XW}$ and $\overrightarrow{XK}$, α=54.7° for $\overrightarrow{XW}$ and $\overrightarrow{\Gamma L}$

Multi directional analysis will give more reliable evidences for the process I (III) through the scattering between Lc and X1 or Xu (X1 and Xu) depending on their photo-reactive energy gaps in UV to visible region. The characteristic conditions are realized at $\theta$=63° and 117° as shown in Fig. 12 when the direction of incident photon matches with Γ-W direction. This induces some peaks to arise in the observed photocurrent as a function of $\theta$.

3-1 Modified approximate expression for the band-to-band scattering

The approximate expression used in this study referred to the equation[26]:

$$R_i cos^2\theta + R_t sin^2\theta \qquad 7.$$

where $R_i$ ($R_t$) is the coefficient calculated from effective mass of each conduction band, $\theta$ the observed current direction. We modified Eq. 7 in order to match with our experiments as functions of $\phi$ and $\theta$ because Eq. 7 can be applied when the relative angle of 90° between related bands. We used vector-crossing angles: Δ=18.44° for XW-XK、α=54.7° for XW-$\Gamma L$ as shown in Fig. 12. These angles were calculated from vector components for the related combinations as indicated in Fig. 6.

Calculation models of electron existence probability in the conduction bands had been proposed for the equivalent inter valley by J. L. Egley and D. Chidambarrao[27] and for the non-equivalent inter valley by S. Dhar et al.[28] Following these preceding works, Koganemaru et al.[26] successfully simulated the mobility modulation induced by strain and indicated Eq. 7 is useful for the band-to-band electron transition through phonon scattering at room temperature.

Another requirement for our study is the suitable expression should be selected for each band depending on their function as a carrier source or reservoir. We estimated photo currents ($J_{KW}$) as a function of $\theta$ in relative quantity by the following modified equation:

$$J_{KW}=\frac{1}{\mu_k}\left(\mu_W \sin^2(\theta-\Delta)+\mu_K \sin^2\theta\right)\cdot\sin\theta \tag{8.}$$

where the coefficient is given as mobility; $\mu_W$ is the effective mass for X1 (X-W direction), $\mu_K$ for Xu (X-K direction), Δ=18.44° the crossing angle of XW-XK as indicated above. Two conduction bands: X1 and Xu can be regarded as a carrier source simultaneously because some of electrons initially occupied at around X ($k$~2.0×π/a $cm^{-1}$) by doping as shown in Fig. 10. 1st and 2nd sine functions express this doping condition and 3rd one reflects the shared factor of incident photon maximized at $\theta$ =90°.

On the other hands, the following equation:

$$J_{WL}(\pm)=\frac{1}{\mu_L}\left(\mu_W \sin^2(\phi-\alpha)\pm C_L\mu_L \cos^2\phi\right)\cdot\cos\phi \tag{9.}$$

where $\mu_L$ is the effective mass for Lc (Γ-L direction), $\mu_W$ for X1 (X-W direction), was applied for the photo currents ($J_{WL}$) in the rotation as a function of $\phi$. The sine function represented XW as a carrier receiver and 1st cosine for Lc as a carrier source realized with UV excitation, expressing the contribution ratio ($C_L$) by the carrier generation process I. The positive/negative sign with $C_L$ was given as a best fitting value in comparison with experiments because Lc can be a carrier source or receiver depending on $\phi$ and excitation wave length. The phase angle of α=54.7° for a vector crossing of XW-$\Gamma L$ and the 2nd cosine expresses Γ-L component shared factor of the incident photon. The mobilities used in Eq. 8 and Eq. 9:

$$\mu_i \propto \frac{\mathrm{DOS}_i}{\left(m_i^*\right)^{3/2}\cdot E_i^2}\quad (i=K,W,L) \tag{10.}$$

are composed of density of state ($\mathrm{DOS}_i$), deformation potential ($E_i$) and effective mass ($m^*_i$), (i=K, W, L). These approximations are originated from the scattering rate (1/τ)[29,30] expressed by

$$\frac{1}{\tau}=\frac{(2m_i^*)^{3/2}E_i^2 k_B T}{2\pi\hbar^4\rho\upsilon_S E_k^{1/2}}\propto const.\times\frac{(m_i^*)^{3/2}\cdot E_i^2}{\mathrm{DOS}\left(=E_k^{1/2}\right)} \tag{11.}$$

at room temperature. We supposed each mobility of related band varies with those factors: DOS, $E_i$ and $m_i^*$, being in proportion to the relaxation time (τ):

$$\tau \propto \frac{\mathrm{DOS}_i}{\left(m_i^*\right)^{3/2}\cdot E_i^2}\quad (i=K,W,L) \tag{12.}$$

is proportional to mobility ($\mu$). DOS here were approximated by the square root of E($k$) for the simple analysis. Modified expressions: $J_{KW}$ ($J_{WL}$) are normalized by $\mu_K$ ($\mu_L$) so as to evaluate the relative fluctuations as a function of $\phi$ ($\theta$) in comparison with experiment. Both of the deformation potentials: $E_i$ in Eq. 10 and $C_L$ in Eq. 9 were given for each conduction band by best fitting with measured photocurrents.

Normalized photocurrents ($J_{KW}$, $J_{WL}$) depend on excitation wavelength since process I for Lc-Lv in UV region and III for X1-Xu (in Visible) band to band transition as shown Fig. 9. Considering the related reciprocal lattice points in Fig. 6, characteristic angles at θ=63° and 117° are predicted for $J_{KW}$ when the incident photon in the direction of Γ-W as mentioned previously. $J_{WL}$ with $\phi$ rotation, in contrast, will fluctuate between $J_{WL}(-)$ to $J_{WL}(+)$ with $C_L \neq 0$ depending on scattering intensity and $C_L=0$ is predicted when any effects from Lc to X1 disappeared.
The scattering effect within X1, Xu and Lc as two variables ($\phi$ and $\theta$) is estimated by the expression:

$$J_L(\phi,\, \theta) = J_{KW} + J_{WL}$$

$$= \frac{1}{\mu_K}\left(\mu_W \sin^2(\theta - \Delta) + \mu_K \sin^2\theta\right) \cdot \sin\theta + \frac{1}{\mu_L}\left(\mu_W \sin^2(\phi - \alpha) \pm C_L \mu_L \cos^2\phi\right) \cdot \cos\phi \qquad 13.$$

Finally, the observed current flow direction (110) combined with $J_L$ in the relation:

$$J_\theta = J_L \cdot \cos^2(\theta - 45°) \qquad 14.$$

was used for the evaluation as a function of $\theta$. The last cosine is defined by the relative arrangement between (110) and the direction of incident photon.

## 3-2 Results and discussion for multi-directional analysis

Figures 13 and 14 show variations in photocurrent as a function of $\phi$ ($\theta$) with excitation of 365 and 405 nm, respectively. Scanning ranges of angle are $\theta$=30° to 130° with $\phi$=0° to 150°, so as to avoiding any mechanical interference through rotation. The photocurrents in the left scale are negative due to reversed photovoltaic signal as measured raw data. This is why we focused on those relative variation for the following discussion.

### 3-2A UV excitation at 365 nm

$J_\theta$ (Eq. 14) gives a better approximation to the measured variance as a function of $\theta$ in comparison with $J_L$ (Eq. 13) under excitation at 365 nm as shown in Fig. 13 (a), indicating the direction of incident photon is an effective factor for this condition. A couple of broad peaks are observed at $\theta$=50° and 110° in Fig. 13 (a), being close to the predicted angles: $\theta$=66°, 117° in the head of Sec. 3. Therefore, Eq. 14 ($J_\theta$) can be applied for the better simulation rather than Eq. 13 ($J_L$), including scattering effects within K-W-L and the direction of incident photon.
$J_{WL}(-)$ (fitting value $C_L$=+0.2) as a function of $\phi$ much closes to the variations in measured photocurrent in the right of Fig. 13. A decline around $\phi$=70° in $J_{WL}(-)$ entirely simulated the measured results for every conditions of $\theta$=30° to 130°. In contrast to this, $J_{WL}(+)$ were not applied for any approximations. These results indicate the negative photo-carriers increased by the scattering from Lc in proportional to "-$C_L$" in $J_{WL}(-)$ under UV excitation.
A best fitting of $J_{WL}(-)$ gave ratios of deformation potential for Xu/X1 =0.5 (in a combination of X-K/X-W) and for Lc/X1=0.3 (ibid. Γ-L/X-W), respectively. An obtained fitting value: $C_L$=+0.2 differs in the polarity with another excitation wave length (see the following section). Considering Eq. 9 with this polarity, the scattering from Lc to XW is the predominant under UV excitation therefore it supports a large sensitivity at $k$~1.3 corresponding to Van Hove singularity: I in Fig. 8, ibid. at ~3.5 eV (UV region) in Fig. 9.

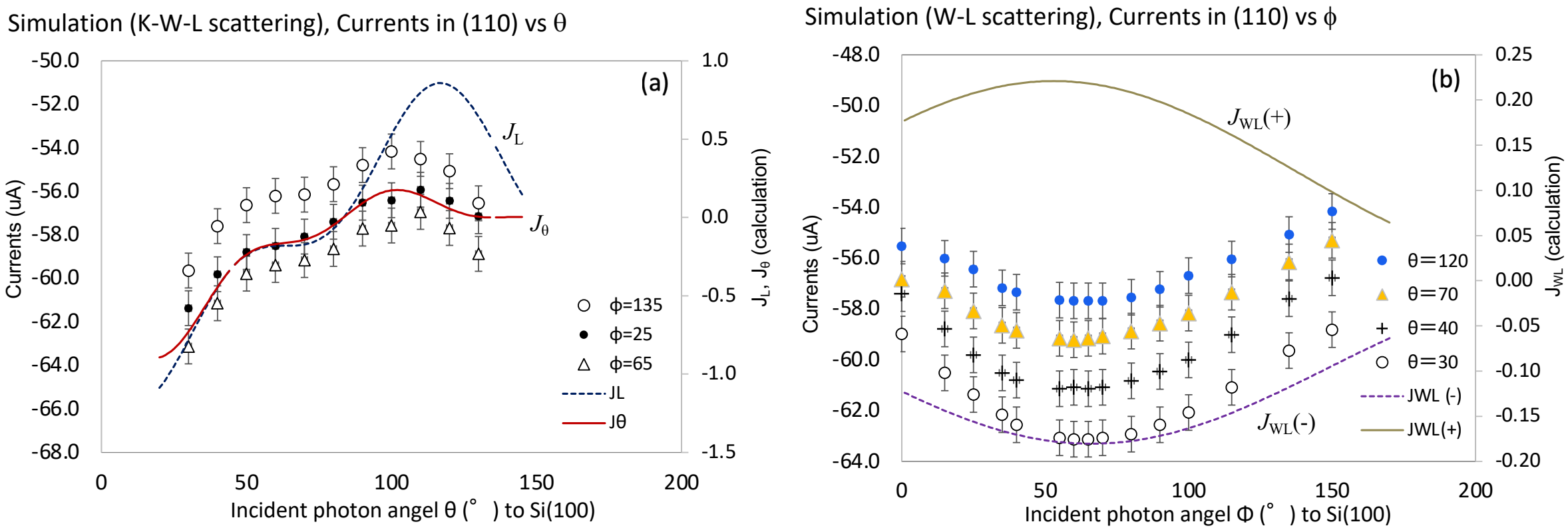

Fig. 13: Directional analysis under the photo excitation (365 nm, 3.397 eV) closed to a maximum DOS at 355 nm (3.494 eV) with $k$=1.313 (see Table 1 and Fig. 7). The error bars due to the back ground fluctuations by the stray light. This insufficient excitation energy of ~100 meV is compensated by the thermal excitation at 300 K (~100 meV). LED as light source ranges in wavelength over the condition at maximum DOS. The broad optical range with thermal excitation gave rise to ~3.7% power efficiency of the LED used for the excitation (see Fig. 18 in Appendix 1). On the other hand, a better quantum efficiency (~10%) was observed at 3.40 eV in Fig. 9 against that lower excitation power.: (a) Photocurrents (circles, triangles and cross) with error bars measured in (110), simulations ($J_L$ and $J_q$) for W-K-L scattering. (b) Simulations ($J_{WL}$) for W-L scattering with a fitting value of $C_L$=+0.2.

A downward peak around $\phi$=70° in Fig. 13 (b) is explained by process I and III introduced in Sec. 2-1. The wave number: $k$~1.3 in the process I is larger than the maximum ($k_z$=0.71) of Γ-L excitation (see the Sec. 3-2B), being satisfied through $\phi$-rotation. This led a simple dependence on $\phi$, where it differs from another excitation wavelength. In contrast to this, a direction of observed current (110) existing in the X-W-K plane (see Fig. 6 and Fig. 12) means the photocurrent reflects the carriers through scattering from Lc to X1 or Xu because the predominant carries are generated in Lc (process I) under excitation at 365 nm. Thus, combination of these factors realizes a downward peak at $\phi$=70°, following the expression $J_{WL}(-)$, being consistent with experiment.

Considering the scattering from X1 to Xu in the process III, the first term in Eq. 9 is supposed to be an effective factor and, consequently, some fluctuation depending on $\theta$ is expected through $\phi$ rotation. The results in Fig. 13 (b), however, are not consistent with this supposition and being independent from $\theta$, thus the contribution by process III is much smaller than that of process I under UV excitation.

Discussion above supports the multi directional sensitivity which is originated from the combination of Van Hove singularity and the phonon scattering from Lc to XW(K) for UV excitation.

3-2B Visible excitation at 405 nm

Fig. 14 (a) shows $J_\theta$ (Eq. 14) well simulates the measured photo current as a function of $\theta$ with a broad peak at 60°~80°. A best fitting by $C_L$=-0.5 in Fig. 14 (a) means carrier transitions between X1 (or Xu) and Lc in the reverse direction of UV excitation in Sec. 3-2A. The broad peaks at $\theta$=66°, 117° simulated as observed for 365 nm excitation in Fig. 13 (a). $J_{WL}(-)$ or $J_{WL}(+)$ in Fig. 14 (b) were selected for the fitting depending on $\theta$, therefore, the carrier supplies from Lc is an effective factor according to conditions.

A couple of characteristic peaks: P1+, P1- were observed at combination angles of ($\phi$, $\theta$) = (120°, 80°) and (90°, 120°) through $\phi$ rotation in Fig. 14 (b). These peaks arisen on $\phi$-rotation curves are explained by the relative configurations of the reciprocal lattice points and the direction of incident photon as follows.

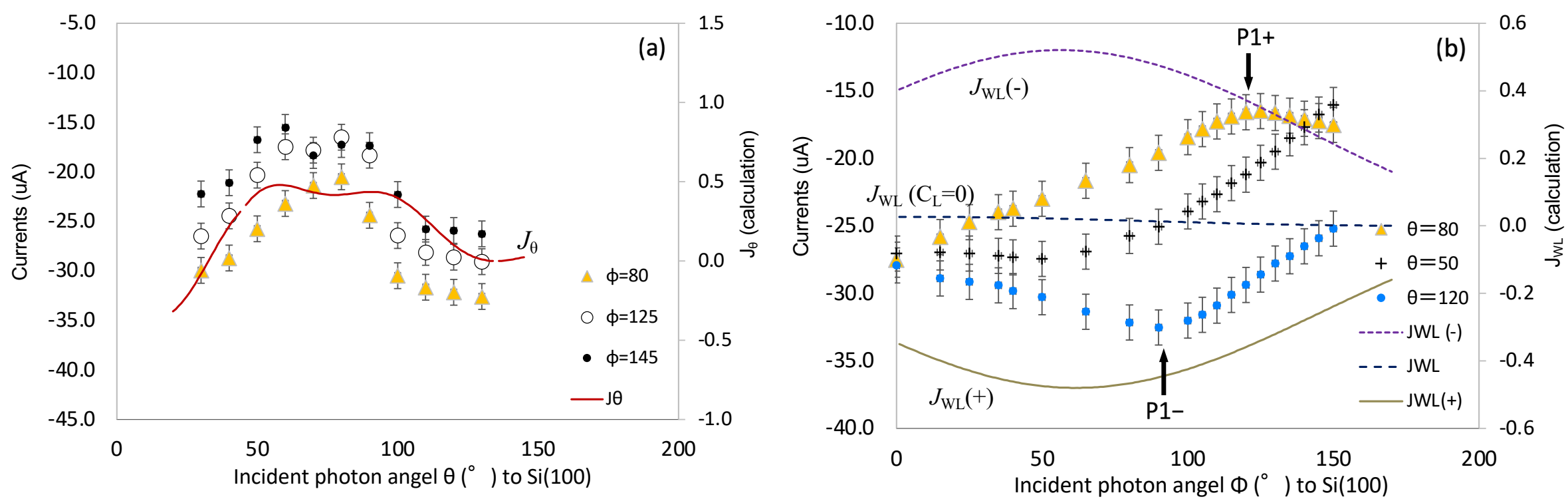

Fig. 14: Directional analysis under photo-excitation of 405 nm (3.06 eV) with error bars. (a) The photo current measured (circles and triangles) as a function of $\theta$ and a curve ($J_\theta$) with a best fitting value: $C_L$= -0.5. A couple of peaks on the curve ($J_\theta$) well simulates the predicted angles in the head of Sec. 3. (b) The measured photocurrent (circles and triangles) and simulation curves: $J_{WL}(+)$ and $J_{WL}(-)$ as a function of $f$. In contrast to the UV excitation (Fig. 13 (b)), positive and negative peaks corresponded to $J_{WL}(-)$ and $J_{WL}(+)$ with the reverse relation.

Fig. 15 illustrates an arrangement of the reciprocal point of K, W, L as a Γ-X rotation axis view at $\phi$=90°. The quantum efficiency is enhanced in the case of matching the dimension of $\overrightarrow{K_Z}$ with the vector component of the incident photon ($\overrightarrow{K_P}$). The miss match direction between $\overrightarrow{K_Z}$ and $\overrightarrow{K_P}$ is supposed to be compensated by two factors: the multi-direction of the metal crystal faces on n-Si (Fig. 2), the phonon scattering inter conduction bands.
The dimension of $\overrightarrow{K_Z}$ are expressed by

$$k_Z(\phi,\ \theta) = \Delta k_Z \sin\phi \sin\theta \qquad 15.$$

where $\Delta k_Z$ = 0.5, 0.75 and 0.71 (in unit of 2π/a $cm^{-1}$) for X-W, X-K and Γ-L, respectively. K, W, Γ(X) are aligned vertically at $\phi$=90° as a rotation axis of Γ-X (Fig. 15).

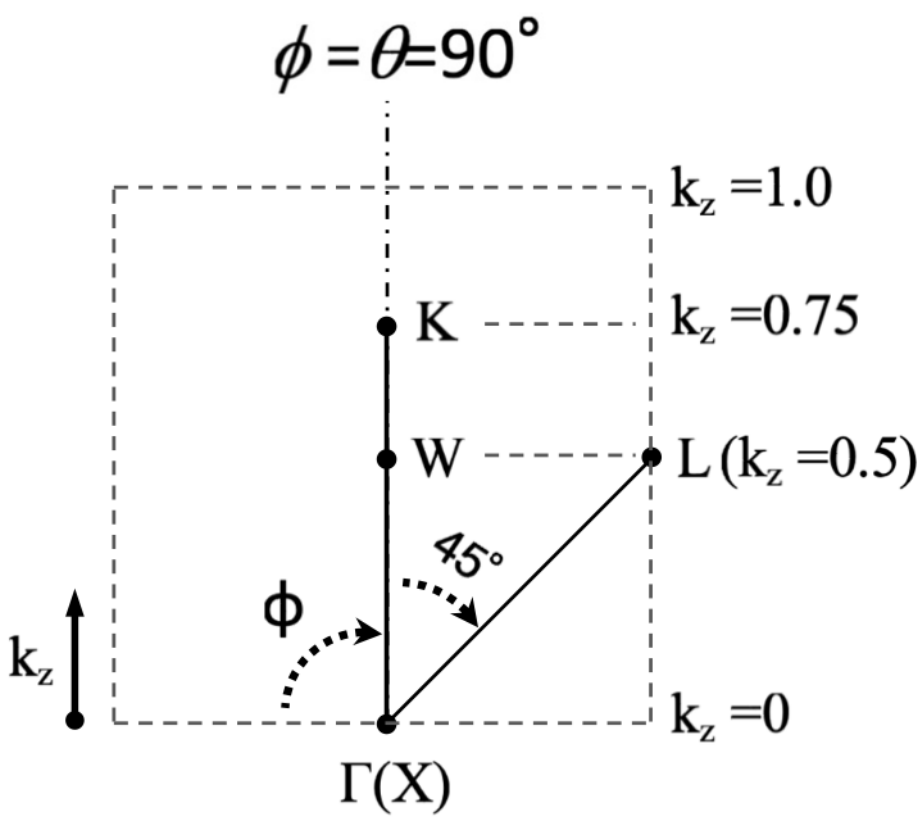

Fig. 15 Relative configurations of the related reciprocal points in a Γ-X view at $\phi$ =$\theta$=90°. The dimensions of $\overrightarrow{K_Z}$ are 0.75 and 0.5 (in unit of 2π/a $cm^{-1}$) at this angle, corresponding to K and W(L), respectively. The radius of scanning circle means $\Delta k_z$ in Eq. 15, e.g., a radius of 0.71 for Γ-L rotation.
An anomaly on excitation efficiency from X1 to Xu will be expected when $k_Z(\phi,\theta)$ closes to the z-component of $\overrightarrow{K_P}$.

This "$k_p$" requires to be larger than $k_z$ because of some scattering loss. The error "$|2k_Z(\phi,\theta) - k_P|$" is available on inspection of transfer efficiency from $k_p$ to $k_z$. Thus, we deduced this error made an exceptionally increase (or decrease) at P1+, P1- and predicted each of the angles to emerge peak(s) by the following relation:

$$I_P = \frac{1}{|2k_Z(\phi,\ \theta) - k_p|} \propto J_{WL} \qquad 16.$$

where $I_p$ is the matching intensity in arb. unit., $k_p$ the z-component of incident photon. This trial function specifies the angles ($\phi$, $\theta$)) in inverse proportion to the error "$|2k_Z(\phi,\theta) - k_P|$". The polarity of $J_{WL}$ was negative (positive) when carriers transferred from (to) Lc (Ip L). (see Appendix 2 for the improved $I_p$ by using $k_X$, $k_Y$ and $k_Z$ with $DOS_{cv}$ of the related energy bands including XK).

Fig. 16 (a) shows $I_p$ for W (Ip W, solid curve) and L (Ip L, points with solid curve) at $\theta$=80°. A positive peak at $\phi$=120° (P1+) closes to a middle peak of W+L as shown in Fig. 16 (b). The optimum fitting with experiment (triangles) was obtained with $k_p$=0.801 for an energy gap of 2.849 eV, listed in Table 1, as shown in Fig. 16 (b). Using Eq. 16, another effect from XK was impossible to give a reasonable fitting for $\theta$=80°. One of the reasons why the energy gap (4.068 eV) of XK is higher than the excitation energy (3.06 eV). This means the inefficient transition occurs through XK by the phonon scatterings. The difference in the angle between the peak at $\phi$ =100° (simulation) and 120° (P1+, triangles) was due to $I_p$ (Eq. 16) used $k_z$ without other factors: $k_x$ ($k_y$), DOS and mobility of each band. Fig. 16 (c) shows the configurations of the related points with a circular guide for eyes ($k_p$=0.801) at ($\phi$, $\theta$) = (120°, 80°). $2k_z$ for W closes well to the circle, and therefore, the factor of Ip W expected a major factor under this condition. A major peak of Ip L at $\phi$~100° joined with double peaks of Ip W then assumed to compose a single peak at 120° in experiment. The positive peak means a decrease in the photo carriers following $J_{WL}$(-) in Fig. 14 (b) and it indicates electrons transfer to Lc as a temporal reservoir.

In contrast to P1+ in Fig. 16, a negative peak at $\phi$ =90° (P1-) was observed for $\theta$=120° in Fig. 14 (b). P1- was assigned to $k_p$=0.62 from the simulation in Fig. 17 (b) and a corresponding energy gap of 2.120 eV was estimated from the energy band diagram in Fig. 5. Fig. 17 (c) shows the relative points configuration for this situation. This downward peak is explained by a smaller intensity of L as shown in Fig. 17 (a) when W factor is joined together. This is consistent with the decrease in L factor because $2k_z$ for L is far from the circle in Fig. 17 (c). The results above ($\theta$=80° and 120°) indicate the electron transfer from/to Lc is impossible to be ignored. The phonon scattering generally assists the inter band transitions thus, some carriers from Lc are indirectly acceptable even if the excitation condition differs from the optimized configuration at Van Hove singularity in UV region.

Although these simulated (W+L) peaks close to the observed angles, they are much broad in experiments (see Fig 14). The following factors should be considered for the improvement: the scattering from/to the related conduction bands (XW, ΓL and XK), the other components of $k_x$, $k_y$ in Ip, DOS of the related bands. This is because the reciprocal points positioned in a twisting relation with each other and the direction of incident photon relatively changed depending on the scanned angle of ($\phi$, $\theta$). Except for the phonon scatterings, improved simulations included 3D (x, y, x) factors with DOS will be presented in Appendix 2.

The characteristic peaks in $\phi$-scanning (Fig.13 (b) and Fig.14 (b)) matched simulations by Eq. 16 and thus, process III introduced in this study is a reasonable carrier generation model.

Following the variance of the factor: $C_L$ in Eq. 9 ($J_{WL}$), positive $C_L$ (+0.2) when the scattering from Lc is dominant under UV (365 nm) excitation (Fig. 13), negative $C_L$ (-0.5) for the carriers going back and forth between Lc and X1 at 405 nm excitation (Fig. 14). Lc as a carrier reservoir assumed to enhance/reduce the negative photo-carriers through the inter-band phonon scatterings.

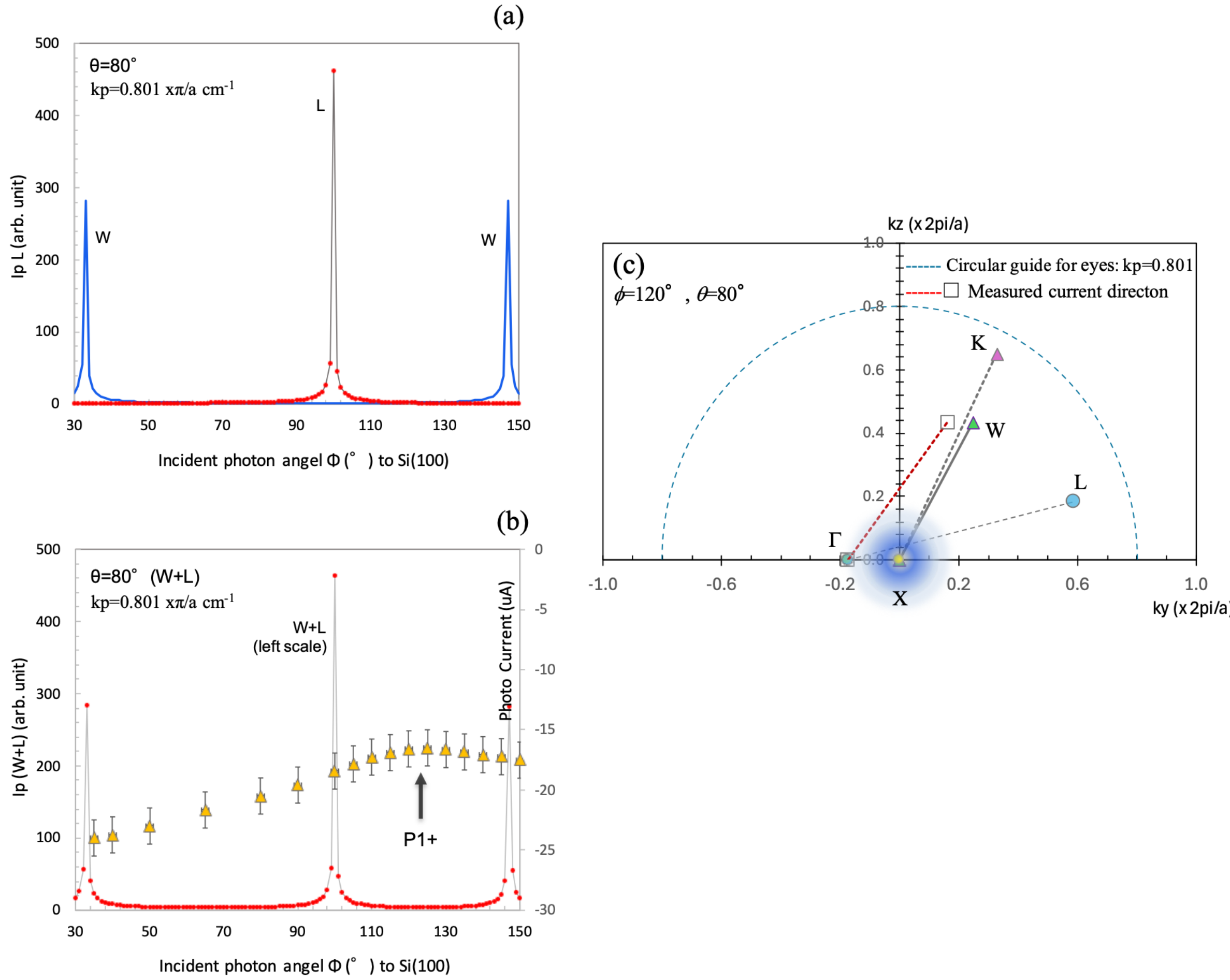

Fig. 16: Simulation of $k_z$-matching intensity ($I_p$, Eq. 16) represented by 1/error ∝ $J_{WL}$ where the error is difference between 2$k_z$ and $k_p$ (=0.801) for $\theta$ =80°, corresponds to an energy gap at 2.849 eV for X1-XW transition in the process III (see Table 1). On the other hand, an excitation energy (3.06 eV) does not exceed 4.068 eV for X1-XK, thus the factor of "XK" is minor in this case (see Appendix 2 for more details). (a) $k_z$ factors in Ip for L (points with curve) and for K (curve), as a function of the incident photon angle (F). Ip L is a dominant peak rather than that of Ip W. (b) Ip (W+L) (points with curve) consisted of W and L factors by adding (W+L). A predicted peak at 100° closes to an observed peak at $\phi$=120° (P1+, triangles) with its polarity for $\theta$=80°. This error is decreased by the simulation coupled with other factors (see Appendix 2). (c) The configuration of the related points at ($\phi$, $\theta$) = (120°, 80°) in view of an incident photon parallel to X coordinate in Fig. 12. A rotation center at X with an electron distribution image and the photo carriers was observed in (110) direction (a red-broken line with square). A circular guide for eyes ($k_p$=0.801) corresponds to a variety of metal crystal faces in Fig. 2 (b).

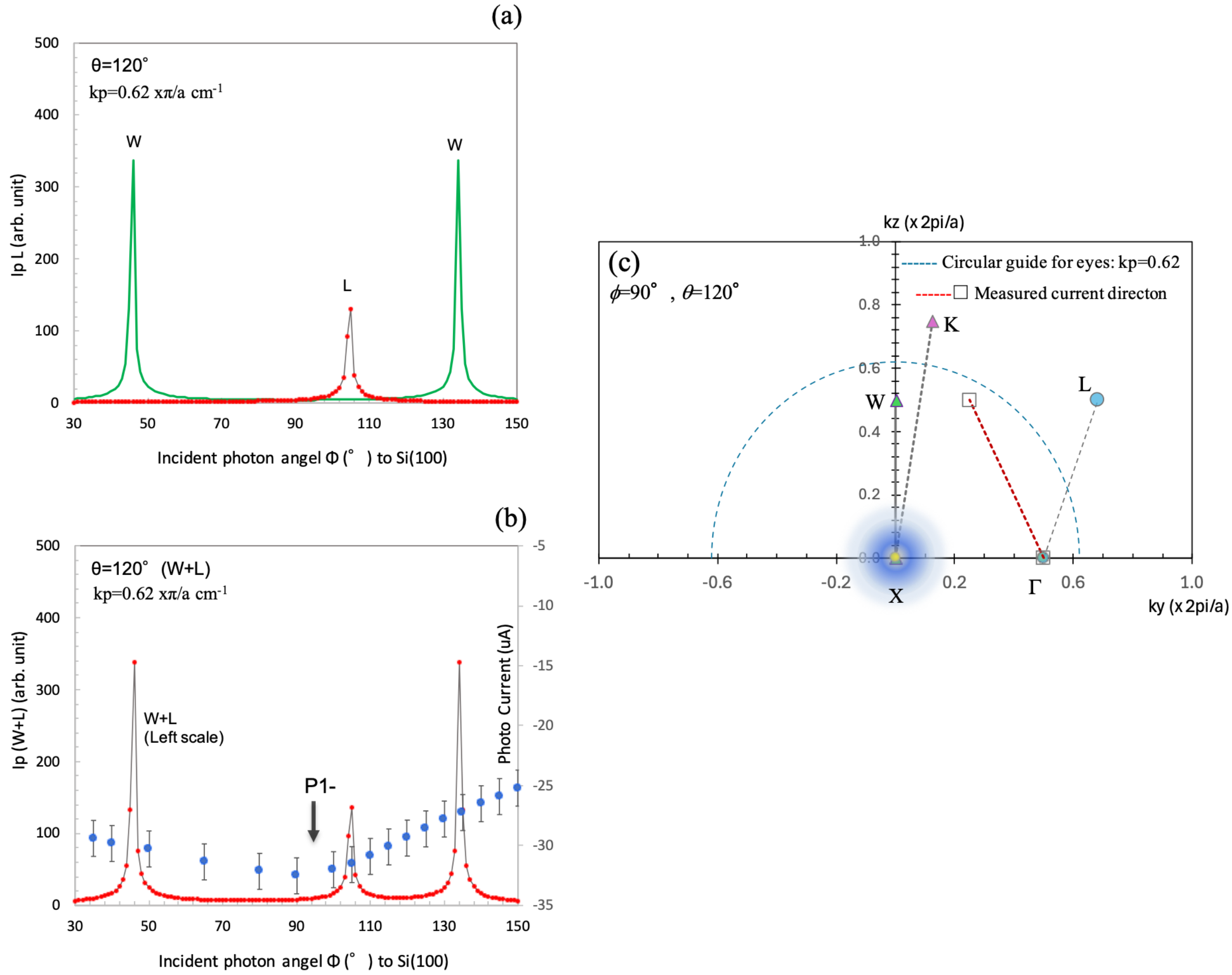

Fig. 17: Simulation of $I_p$ ($k_p$=0.62) for $\theta$=120°, corresponds to an energy gap at 2.120 eV for X1-XW transition in the process III (see Table 1). (a) $k_z$ factors in Ip for L (points with curve) and for W (curve), as a function of the incident photon angle ($\phi$). (b) $I_p$ W+L (points with curve) satisfactorily simulated the observed peak (P1-) at ($\phi$, $\theta$) = (90°, 120°) when L and W factors are joined. (c) The configuration of the related points in view of the incident-photon direction. Γ is shifts to +$k_y$ direction by using a rotation center at X with an electron distribution image. The measured photo carriers (a red-broken line with square) in (110) direction.

**Conclusion**

The proposed process (I) against the inter-band: X1 and Xu was consistent with measured quantum efficiencies, when the transition probability was calculated from the mobility multiplied by the coupling DOS for visible region (1.0 to 2.7 eV).
The doping process (IV), difference in the carrier density between X1 and Xu, well simulated the optical responses for NIR (0.6 to 1.0 eV). Filling a zero gap between these bands by doping with Nd=$1x10^{18}$ ($1/cm^3$), the theoretical threshold of sensitivity agreed with a sharp rise in the sensitivity observed at 0.6 eV.
Van-Hove singularity at reciprocal lattice point: L promotes the UV response above 3.0 eV.
The combination of processes above dramatically released the spectral range from the limitation by a Si-band gap, resulted in the wide optical-response found on the metal/n-Si system.
The multi-directional analysis of observed photo current unveiled multiple carrier formation processes: indirect transition from X1 to Xu, inter-band (X1, Xu and Lc) scatterings and the direct transition due to Van Hove singularity at L point are in cooperated with each other. Acting as a carrier supplier under UV excitation, the conduction band: Lc assisted the spatial photo detection independence from the in-plane carrier formation between X1 and Xu.
This is because the multi-crystal faces of metal on Si, as the optical wave guide to Si surface, are supposed to increase photo carriers against 3-dimensional space.
The measured photo current as a function of excitation angles ($\phi$ and $\theta$) were explained by the modified scattering model. Characteristic angle at each peak in there was consistent with estimations from their relative configuration of the reciprocal lattice points, the shared factor of incident photon, depending on excitation wave length.
The origin of multi-directional response has been explained by a spatial detection by Lc with Γ-L (111) direction cooperating with in-plane carrier formation between X1 and Xu. The carrier scatterings inter conduction bands assist the carrier formation processes to synchronize each other.

**Prospects and Future Issues**

Responsible range:
UV-A~350 nm to NIR~3000 nm is possible in practical uses (to be opened elsewhere).
Factors to be optimized:
doping density, arrangement of electrodes, size scale and forms of top metal structures. Multi-directional response will be improved by the suitable structures of top metal.
Another material:
The carrier formation processes discussed in this study can be applied for Ge because a zero-gap at around a reciprocal lattice point (X) similarly existed in those crystal. Furthermore, the anisotropic crystal faces naturally formed by surface etching process for Ge.[31]
Advanced measurement:
Demonstrating estimation of deformation potentials for Xu/X1 and Lc/X1, the method will be applied for other magnetic materials with polarized photo excitation.

**Appendix 1**
An explanation for the lower quantum efficiency against a sharp DOS peak at 365 nm.

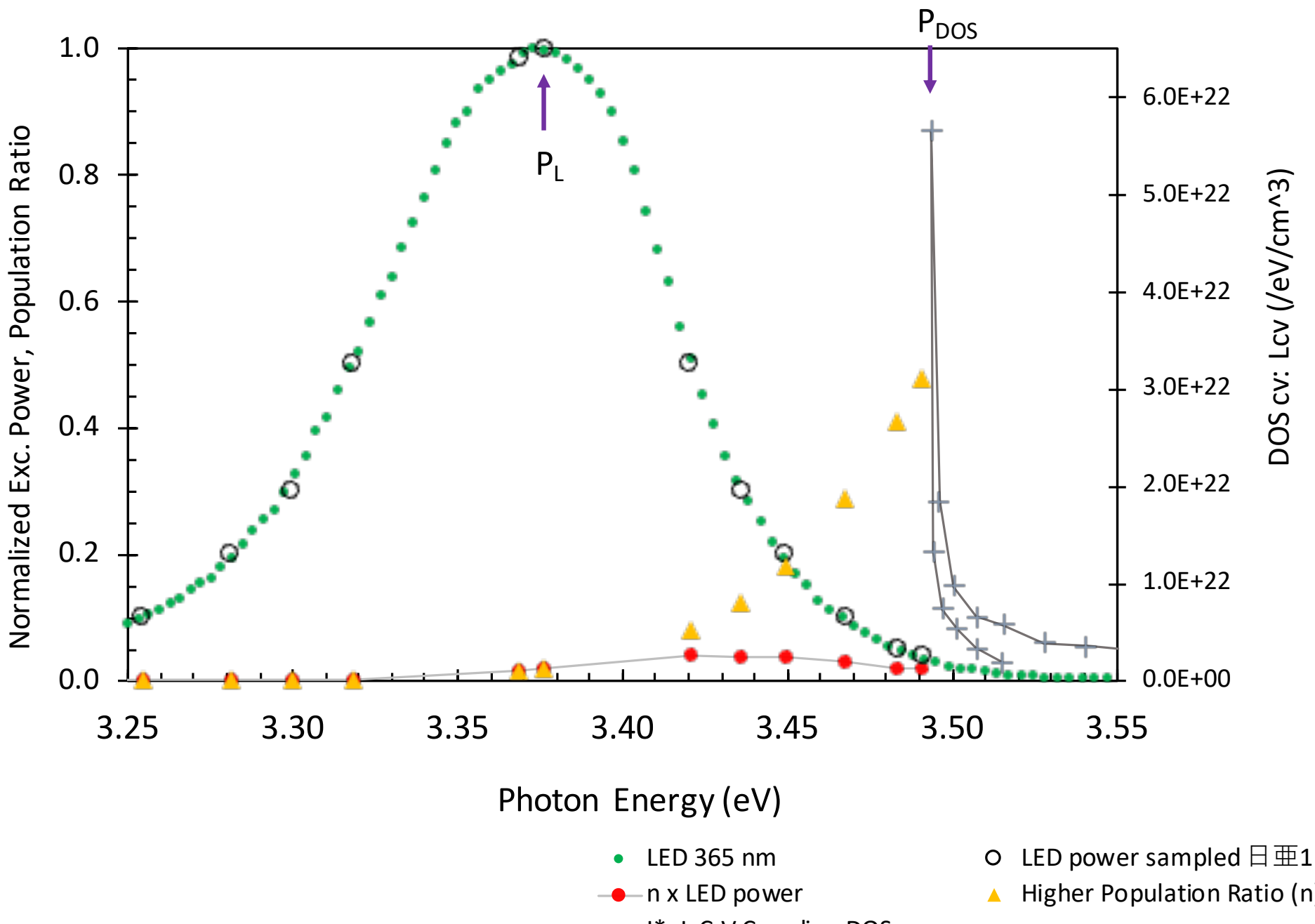

Fig. 18 LED power (closed circles with some of opened: ○), $DOS_{CV}$ around L (crosses with line), the population ratio in the higher state at 3.49 eV (closed triangles) and the available excitation power (closed circles with line) as a function of photon energy. The vertical scale of DOS is displayed in the right and the others in the left scale. LED (365 nm) as an excitation light source with its peak power at ~3.38 eV ($P_L$) is insufficient energy to a maximum $DOS_{CV}$ at 3.49 eV ($P_{DOS}$). This energy difference ~100 meV is closed to the lattice temperature at 300 K.

**Precondition:** The mismatch (=$\Delta E \lesssim$100 meV) in the excitation energy to $P_{DOS}$ is compensated by the thermal excitation to $P_{DOS}$ in Lc. The quantitative estimations are as follows:
The photo excitation for the electrons in Lv (valence band at L) will make electrons transfer to the excited state with some broad optical range of LED. The population of this temporary state will be changed by the external perturbations and some of them will be thermally excited to a higher energy state (around $P_{DOS}$). The excitation rate as a function of ΔE will be estimated by the following population ratio of the higher (n) to lower ($n_0$) (a part of the grand canonical ensemble)

$$\text{Population Ratio (higher/lower)}: \frac{n}{n_0} = \exp\left(-\frac{\Delta E}{kT_e}\right) \qquad \text{A1}$$

where *DE* is the energy difference between the photo-excited state and $P_{DOS}$.
This gives the occupation rate in the higher state:

$$\frac{n}{n_0 + n} = 1 - \frac{n_0}{n_0 + n} = 1 - \left(1 + \exp\left(-\frac{\Delta E}{kT_e}\right)\right)^{-1} \qquad \text{A2}$$

where $T_e$: electron temperature (K) derived by

$$T_e - T = \frac{2q\tau\mu}{3k} E_X^2 \left(1 + \exp\left(-\frac{\Delta E}{kT_e}\right)\right)^{-1} \quad \text{A3}$$

at $T$=300 K (lattice temperature), where $E_x$ (=3 kV/cm) is the intrinsic electric field at electrode formerly given by I-V curve in Ref. 1, $q$ the electron charge, $t$ the relaxation time for the carrier scattering (=1x10$^{-12}$ s), $m$ the carrier mobility in Lc and $k$ the Boltzmann constant. The calculated higher population ratio: n (closed triangles) increased as its excitation energy approaches to $P_{DOS}$, i.e., $DE$ closes to 0, e.g. 18% of the photo-excited carriers exist in the level at $P_{DOS}$ for the photon energy of 3.45 eV with $\Delta E$=44.3 meV. The LED power decreased against this approach. Thus, ~3.7% of original LED power is effectively available for the excitation to $P_{DOS}$ as indicated: “n x LED power” (● with lines) in Fig. 18.

**Appendix 2.**
**Improved matching intensity ($I_p$')**
The better accuracy of simulated photocurrents ($J_{WL}$) will be expected if $I_p$ (Eq. 16) includes 3D components of the shared incident photon. Thus, the sum total: $\sum_i \frac{1}{|2k_i(\phi,\theta)-k_p|}(i = x, y, z)$ will give correctly the characteristic angles ($\phi$) corresponding to each peak. Furthermore, DOScv for the related energy band is considered as another gain factor composing the peak intensity. Therefore, the following $I_p$' was used as the improved matching intensity in proportional to the photo current:

$$I'_{\mathrm{P}} = \sum_i \frac{1}{|2k_i(\phi,\, \theta) - k_p|} \times \mathrm{cR} \times n\mathrm{DOS} \propto J_{WL}\ (i = x, y, z) \qquad \text{A4}$$

where $k_i$ is the wavenumber in the lower band (under photo-excitation), $k_p$ the $i$-component of incident photon, nDOS the scale factor to a minimum ($2.15 \times 10^{21}$) in Lcv as listed in Table A1, cR the contribution rate in Table A2.

Table A1 The list of conditions used in the simulation. "nDOS" for the related transitions which were calculated by Eq. 4 between the lower band (X1) and higher XW (XK) and Lcv. $K_p$ (0.80 and 0.63) were obtained from those best fittings for $\theta$=80° and 120°, respectively. (see Fig. 20 and Fig. 21).

| $k_P$ (xπ/a $cm^{-1}$) | nDOS (x $2.15 \times 10^{21}$ /eV/$cm^3$) | | | $\theta$ (deg.) |
|---|---|---|---|---|
| | X1-XW | X1-XK | Lcv | |
| 0.80 | 0.28 | 0.27 | 1.21 | 80° |
| 0.63 | 0.18 | 0.16 | 1.39 | 120° |

Table A2 The energy gaps (Eg) between X1-XK, X1-XW and Lcv with those contribution rate (cR) for the transitions obtained from those best fittings in Fig. 20 and Fig. 21.

| $k_P$ (xπ/a $cm^{-1}$) | X1-XW | | X1-XK | | Lcv | |
|---|---|---|---|---|---|---|
| | Eg (eV) | cR | Eg (eV) | cR | Eg (eV) | cR |
| 0.80 | 2.846 | 1.00 | 4.063 | 0.50 | 3.574 | 0.30 |
| 0.63 | 2.162 | 1.00 | 3.060 | 0.75 | 3.610 | 0.10 |

Another effect on the photo current is due to the relation between the energy gaps (Eg in table A2) and an excitation photon energy ($E_p$) at 3.06 eV as shown in Fig. 19 (b).
The contribution rate (cR) for the transitions were obtained by those best fittings in $\phi$ scanning. A lower contribution (cR=50%) is denoted at $k_p$=0.80 for XK since $E_p$ (3.06 eV) does not exceed the Eg (4.063 eV). Similarly, the smaller rate (<30%) for Lcv originated from the higher energy gaps (~3.6 eV) of Lcv. These factors give rise to the decrease in the photocurrents through the multiple scattering processes.

We used $k$p and cR as fitting parameters in these simulations to compare with measured results in Fig. 20 and Fig. 21. It has been confirmed that the simulation fairly closes to the variation in photo-currents when the peaks are approximated with a broken curve. If the broadening features are ignored in $\phi$–scanning, the angles to found peaks with their intensities can be predicted by the simple model: $I_p$', and therefore, the incident light are shared its power with each conduction bands according to Eq. A4. This supports the multi-bands (XW, XK and Lc) activated as carrier source/reservoir through the scatterings, process I and III with a zero-gap at X, being consistent with the wide optical/directional responses demonstrated by our samples.

On the other hand, each peak should have some broadening factors by scattering processes so as to improve the precision of simulations. Thus far, an updated version of $I_p$', i.e., two factors ($k$p and cR) combined with the scattering factors is under development.

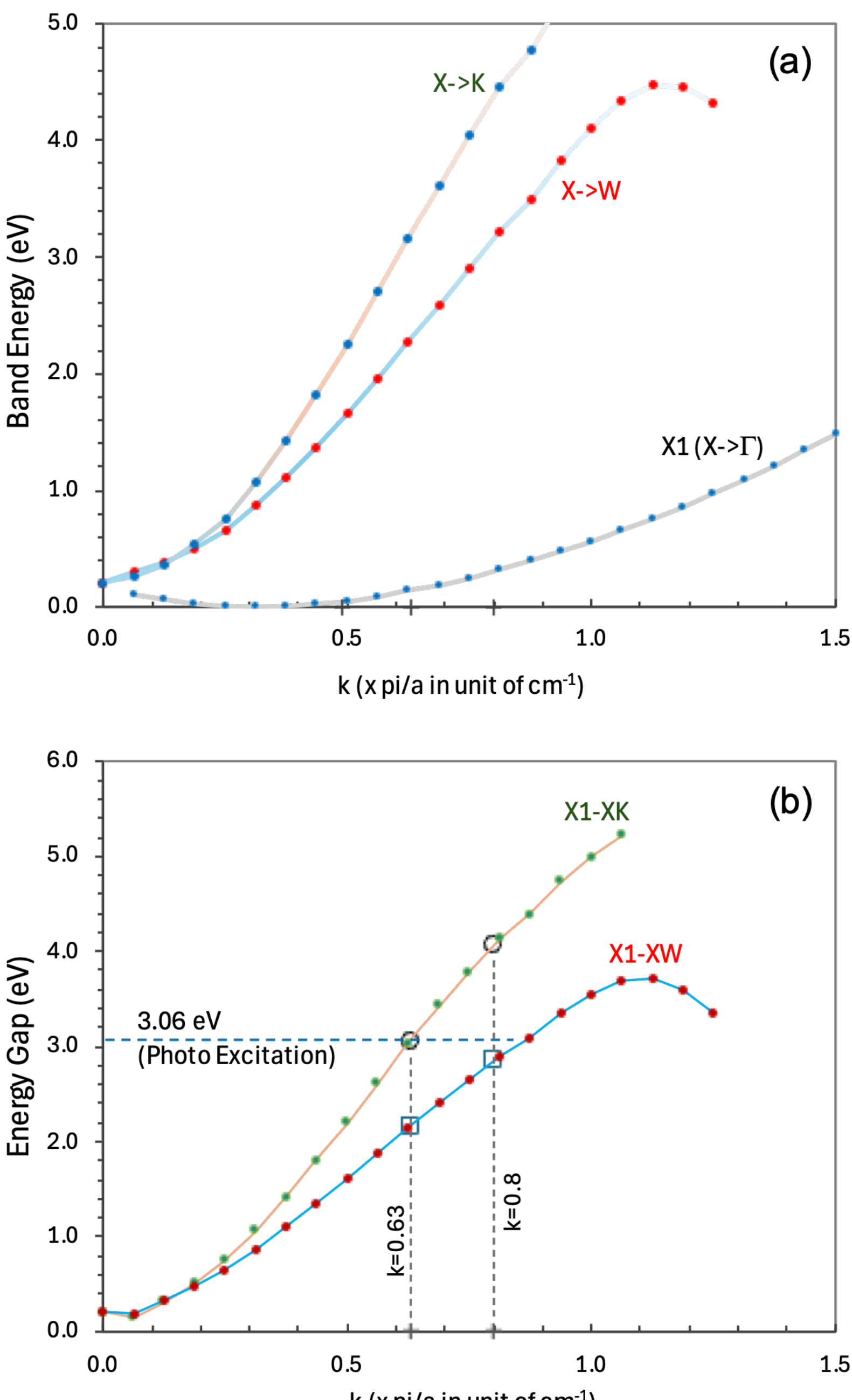

Fig. 19 Band energy diagram and band gaps as a function of wavenumber (k) for the related transitions. (a) Dispersion curves of conduction band (X->W, X->K and X1) are displayed with the horizontal axis as an origin (k=0) at X, being different viewpoint from Fig. 5. This is because photocarriers originated from the zero-gap at X point (see Sec. 2-3 and 2-4). (b) The energy gaps at $k_p$=0.63 and 0.8 on X1-XK (open circles) are more than the photo excitation energy (3.06 eV), thus, the scattering is the major transition process corresponded to the lower cR in Table A2. In contrast to this, the photo absorption on X1-XW (open squares) is promoted prior to scattering processes because the photo excitation energy excesses two energy gaps: 2.162 (2.846) resulted in cR=100%.

Configuration 1: ($k_p$=0.80, $\theta$=80° as revised Fig. 16)

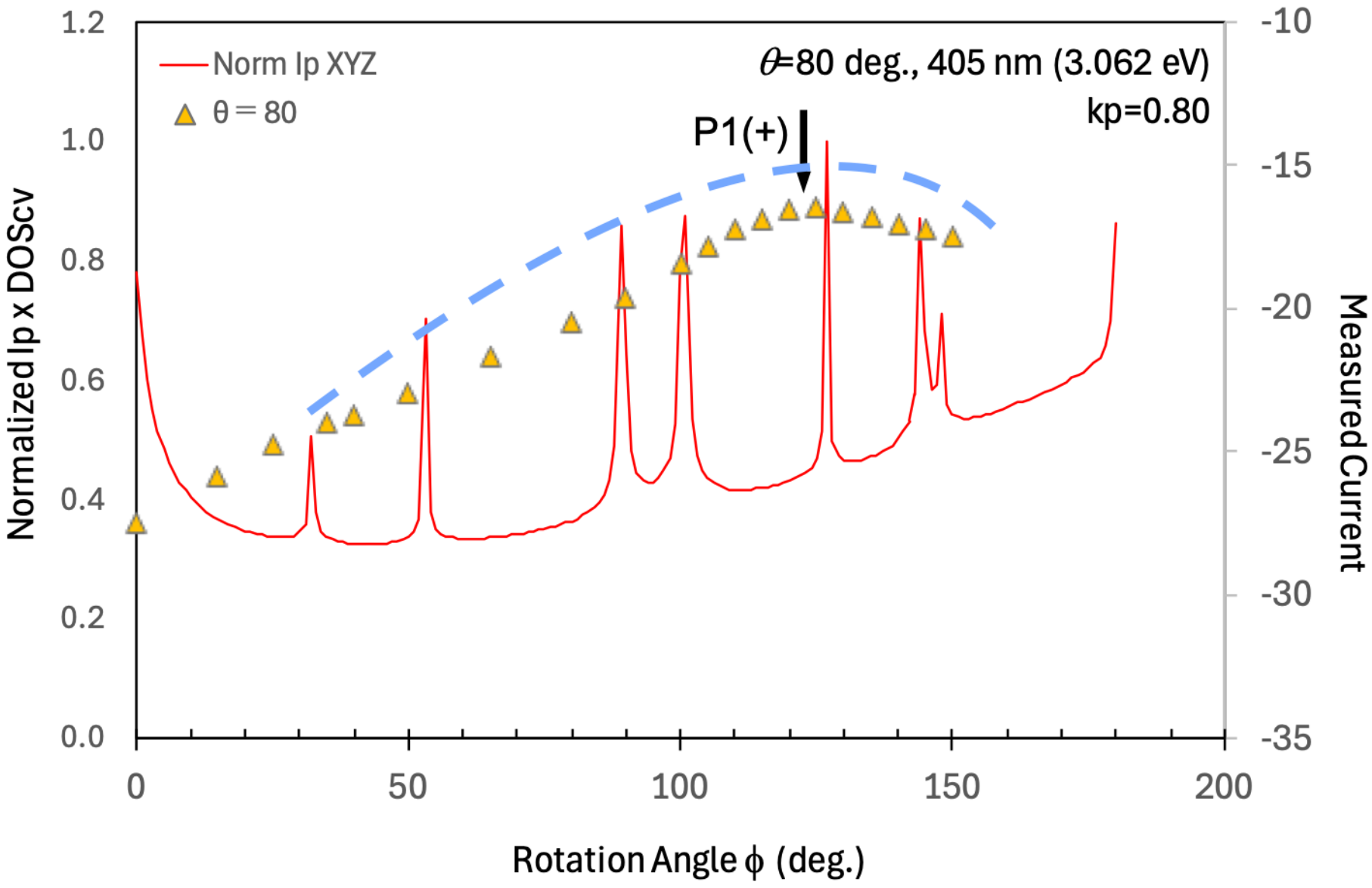

Fig. 20. Simulation (curve) of $I_p'$ multiplied by nDOS and photocurrent (closed triangles) at $\theta$=80° . A main peak at $\phi$=120° (P1+) closes to a largest peak in the simulation. The variation in the peaks composed by kx, ky and kz (broken curve) well simulated the upward feature of the experiment.

Configuration 2: ($k_p$=0.63 for $\theta$=120° as revised Fig. 17)

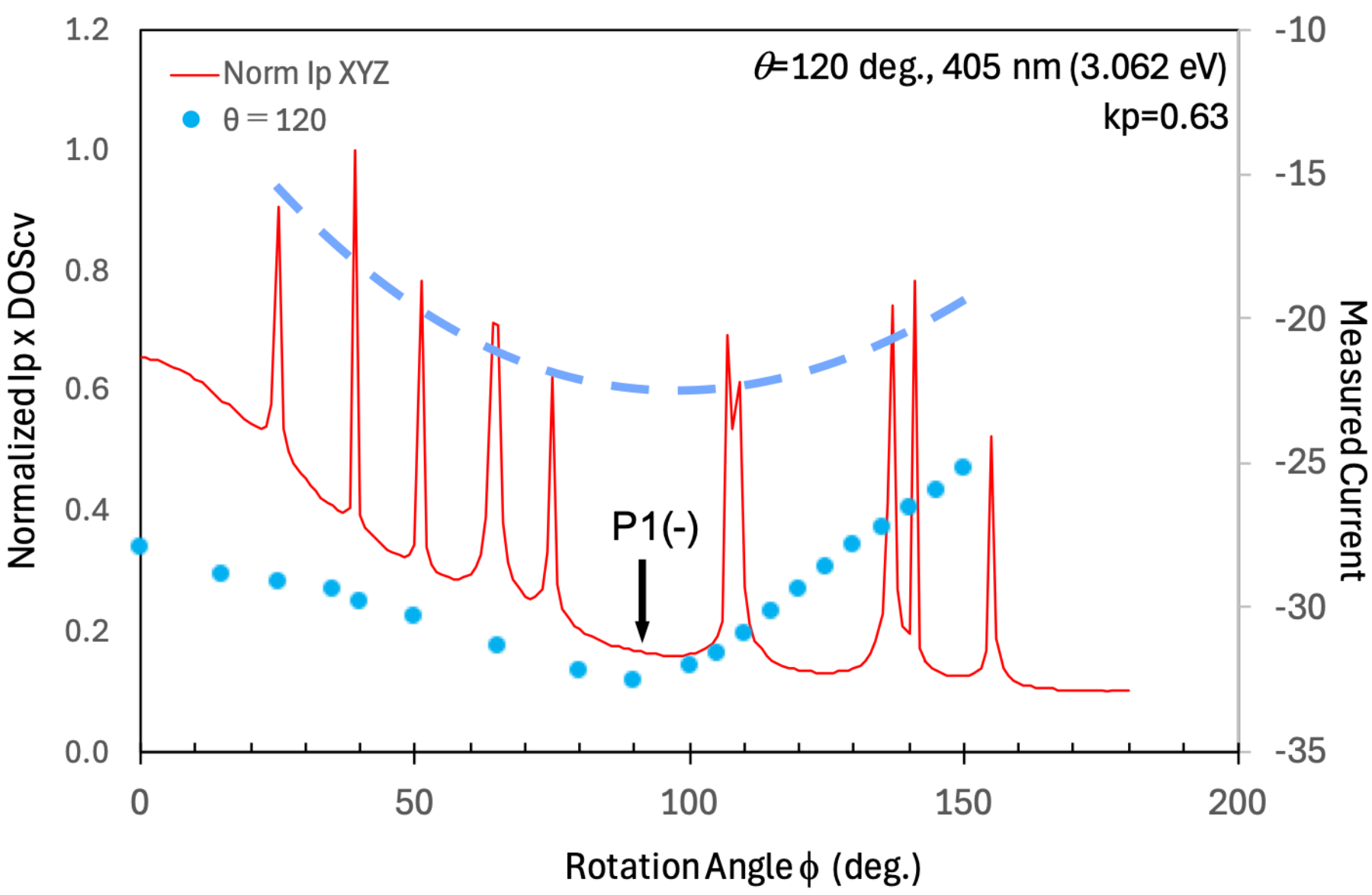

Fig. 21. Simulation (curve) of $I_p'$ multiplied by nDOS and photocurrent (closed circles) at $\theta$=120° . A downward peak at $\phi$=90° (P1-) closes to a bottom in the simulations (broken curve) and the peak-height variation matches with each other.

List of the corrections of 1st version on 24 May 2025 : https://arxiv.org/abs/2505.18678

1. Quantum efficiency substituted for sensitivity of the vertical unit in Fig. 2 to be more precisely plots as a function of excitation photon energy. Fig. 8, 9, and 10 (c) were improved in this revision. In particular, the better approximations, calculation closed to experiment, for 1.5 to 2.7 eV in Fig. 9.
2. The measured data plotted with error bars.
3. Revisions to Sec. 3-1 for more concise expression of calculation processes. Photocurrent formula (Eq. 5) was introduced and mentioned the relation with the other factors: DOScv (Eq. 4), $\mu$ (Eq. 6).
4. Correction of the horizontal axis in Fig. 5, 8 and 9 with the name of the reciprocal lattice points.
5. Displayed a peak of process I in Fig. 7 (b) with its maximum scale (DOScv) at $6.02 \times 10^{22}$.
6. Removed 3-2C (photo excitation at 650 nm) for the simplified contents and avoid being lengthy.
7. Correction of the denominator in the matching intensity ($I_p$: Eq. 16): "$2k_z$" is used as a replacement for "$k_z$" in the previous version.
8. Fig 16 and 17 replaced Fig. 14c in the previous version followed by the above 7th correction.
9. Newly displayed the configuration of the related points in Fig. 16 (c) and 17 (c).
10. Appendix 2 as a supplementary explanations for the incomplete fitting in Fig. 16 (b) and 17 (b). The matching intensity expanded to 3D k-factors with DOScv improved their fitting accuracy.
11. Ref. 6 added as a review works of photo response in crystal Silicon.